\documentclass[usenatbib]{mn2e}
\usepackage{amsmath,amssymb,graphicx,multirow,aas_macros}

\begin{document}
\title{Simulating Subhalos at High Redshift: Merger Rates, Counts, and Types}
\author[Wetzel, Cohn \& White]{Andrew R. Wetzel${}^{1}$, J.D. Cohn${}^{2}$, 
Martin White${}^{1,3}$\\
$^{1}$Department of Astronomy, University of California, Berkeley, CA 94720, 
USA\\
$^{2}$Space Sciences Laboratory, University of California, Berkeley, CA 94720, 
USA\\
$^{3}$Department of Physics, University of California, Berkeley, CA 94720, USA
} 

\date{October 2008}

\pagerange{\pageref{firstpage}--\pageref{lastpage}} \pubyear{2008}

\maketitle

\label{firstpage}

%% ABSTRACT %%%%%%%%%%%%%%%%%%%%%%%%%%%%%%%%%%%%%%%%%%%%%%%%%%%%%%%%%%%%%%%%%%%%
\begin{abstract}
Galaxies are believed to be in one-to-one correspondence with simulated dark 
matter subhalos.
We use high-resolution N-body simulations of cosmological volumes to calculate 
the statistical properties of subhalo (galaxy) major mergers at high redshift 
($z=0.6-5$).
We measure the evolution of the galaxy merger rate, finding that it is much 
shallower than the merger rate of dark matter host halos at $z>2.5$, but 
roughly parallels that of halos at $z<1.6$.
We also track the detailed merger histories of individual galaxies and measure 
the likelihood of multiple mergers per halo or subhalo.
We examine satellite merger statistics in detail: $15\%-35\%$ of all recently 
merged galaxies are satellites and satellites are twice as likely as centrals 
to have had a recent major merger.
Finally, we show how the differing evolution of the merger rates of halos and 
galaxies leads to the evolution of the average satellite occupation per halo, 
noting that for a fixed halo mass, the satellite halo occupation peaks at 
$z \sim 2.5$.
\end{abstract}


%% INTRODUCTION %%%%%%%%%%%%%%%%%%%%%%%%%%%%%%%%%%%%%%%%%%%%%%%%%%%%%%%%%%%%%%%%
\section{Introduction}

Mergers are key in the hierarchical growth of structure, and major galaxy 
mergers (referred to as mergers henceforth) are thought to play a crucial role 
in galaxy evolution.
Specifically, they are expected to trigger quasar activity 
\citep[e.g.,][]{Car90}, starbursts \citep[e.g.,][]{BarHer91,Nog91}, and 
morphological changes \citep[e.g.,][]{TooToo72}, and they are thought to be 
related to Lyman Break Galaxies (LBGs), submillimeter galaxies (SMGs), and 
ultra-luminous infrared galaxies (ULIRGs) 
\citep[see reviews by][respectively]{Gia02,Bla02,SanMir96}.
Observational samples of such objects at $z \gtrsim 1$ are now becoming large 
enough to allow for statistical analyses of their counts \citep[e.g.,][]{Ste03,
Ouc04,Cop06,Yos06,Gaw07,Gen08,McL08,PatAtf08,Tac08,YamYagGot08}.

Understanding galaxy mergers and their connection to these observables requires 
comparison with theoretical predictions.
In simulations, galaxies are identified with subhalos, the substructures of 
dark matter halos \citep[e.g.][]{Ghi98,Moo99,KlyGotKra99,Ghi00}.\footnote{
A subhalo can comprise an entire halo if there are no other subhalos within 
the halo, see \S\ref{sec:numerical}.}
Simulations are now becoming sufficiently high in resolution and large in 
volume to provide statistically significant samples \citep[e.g.][]{DeL04,Die04,
GaoWhiJen04,Ree05}.
This high mass and force resolution is necessary to track bound subhalos 
throughout their orbit in the host halo and avoid artificial numerical 
disruption.
This is particularly important for tracking the orbits of galaxies, which are 
expected to reside in the dense inner core of subhalos and to be more stable 
to mass stripping than dark matter because of dissipative gas dynamics.
The correspondence of galaxies with subhalos has been successful in reproducing 
galaxy counts and clustering in a wide array of measurements 
\citep[e.g.,][]{Spr01,Spr05,ZenBerBul05,Bow06,ConWecKra06,ValOst06,WanKauDeL06}.
Henceforth, we will use the term galaxy and subhalo interchangeably.

A subhalo forms when two halos collide and a remnant of the smaller halo 
persists within the larger final halo.
Thus, subhalo merger rates are sometimes inferred from halo merger 
rates\footnote{
Halo merger counts and rates have been studied in a vast literature, both 
estimated analytically \citep[e.g.,][]{KauWhi93,LacCol93,PerMil99,BenKamHas05,
ZhaFakMa08} and measured in simulations \citep[e.g.,][]{LacCol94,Tor98,Som00,
CohBagWhi01,GotKlyKra01,CohWhi05,Li07,CohWhi08,FakMa08,SteBulWec08}.}
or subhalo distributions, using a dynamical friction model to estimate the 
infall time of satellite galaxies to their halo's central galaxy \citep[several 
of these methods are compared in][]{HopHerCox08}.
However, a detailed understanding of galaxy mergers requires a sufficiently 
high-resolution simulation that can track the evolution and coalescence of 
subhalos directly.

Here we use high resolution dark matter simulations to examine subhalo merger
rates, counts, and types, their mass and redshift dependence, and their 
relation to their host halos.
Under the assumption that galaxies populate the centres of dark matter subhalo 
potential wells, our subhalos are expected to harbor massive galaxies 
($L \gtrsim L_*$).
Although our subhalo mass assignment is motivated by semi-analytic arguments, 
our results are independent of any specific semi-analytic modeling prescription.

Previous work on subhalo mergers includes both dark matter only simulations 
\citep{Kol00,Spr01,DeL04,TayBab05,BerBulBar06,WanKau08,Mat08} and hydrodynamic 
simulations \citep{Mur02,TorMosYos04,Mal06,ThaScaCou06,Sim08}.
Many of these earlier studies concern subhalo mergers within a single object 
(such as the Milky Way or a galaxy cluster), others use lower resolution.
We look at relatively large simulation volumes (100 and 250 $h^{-1}$~Mpc boxes) 
with high spatial and temporal resolution.
In addition, many previous works focus on subhalo mass loss and survival rate,
while our main interest here is the population of resulting merged subhalos 
itself \citep[most similar to the works of][]{Mal06,GuoWhi08,Sim08,Ang08}.
We characterize subhalo merger properties in detail, including satellite 
mergers, at high redshift ($z=0.6-5$), during the peak of merger activity.
We also investigate the satellite halo occupation (number of satellites per 
halo), and its evolution as shaped by the relative merger rates of subhalos and 
halos.
The satellite halo occupation is a key element in the halo model \citep{Sel00,
PeaSmi00,BerWei02,CooShe02}, a framework which describes large-scale structure 
in terms of host dark matter halos.

In \S\ref{sec:numerical} we describe the simulations, our definitions of halos 
and subhalos, their infall mass, and the evolution of the satellite fraction.
In \S\ref{sec:rates} we define our merger criteria and examine subhalo merger 
rates and their relation to halo merger rates.
In \S\ref{sec:types} we explore the relative contributions of satellites and 
centrals to the merger population, both in terms of parents and merged children.
\S\ref{sec:counts} gives the distribution of the number of mergers within a 
given look-back time and the fraction of halos that host subhalo mergers.
In \S\ref{sec:hodevol} we show how the difference in halo and subhalo merger 
rates contributes to the evolution of the satellite halo occupation.
We summarize and discuss in \S\ref{sec:discussion}.
The Appendices compare subhalo infall mass to infall maximum circular velocity 
and its evolution with redshift, and give our satellite subhalo mass function.
The halo occupation and clustering of high redshift galaxy mergers is treated 
in \citet{WetCohWhi09b}.

%% NUMERICAL %%%%%%%%%%%%%%%%%%%%%%%%%%%%%%%%%%%%%%%%%%%%%%%%%%%%%%%%%%%%%%%%%%%
\section{Numerical Techniques} \label{sec:numerical}

\subsection{Simulations}

We use two dark matter only N-body simulations of $800^3$ and $1024^3$ 
particles in a periodic cube with side lengths $100 h^{-1}$~Mpc and 
$250 h^{-1}$~Mpc, respectively.
For our $\Lambda$CDM cosmology ($\Omega_m=0.25$, $\Omega_\Lambda=0.75$, 
$h=0.72$, $n=0.97$ and $\sigma_8=0.8$), in agreement with a wide array of 
observations \citep{COBE,Teg06,ACBAR,WMAP5}, this results in particle masses of 
$1.4 \times 10^8 h^{-1}M_\odot$ ($1.1 \times 10^9 h^{-1}M_\odot$) and a Plummer 
equivalent smoothing of $4 h^{-1}$~kpc ($9 h^{-1}$~kpc) for the smaller 
(larger) simulation.  
The initial conditions were generated at $z=200$ using the Zel'dovich 
approximation applied to a regular Cartesian grid of particles and then evolved 
using the {\sl TreePM\/} code described in \citet{TreePM} \citep[for a  
comparison with other codes see][]{Hei07,Evr08}.
Outputs were spaced every $50$~Myr ($\sim 100$~Myr) for the smaller (larger) 
simulation, from $z \sim 5$ to $2.5$.
Additional outputs from the smaller simulation were retained at lower redshift, 
spaced every $\sim 200$~Myr down to $z=0.6$, below which we no longer fairly 
sample a cosmological volume.
For mergers, we restrict these later outputs to $z<1.6$ based on convergence 
tests of the merger rates (we lack sufficient output time resolution in the 
intervening redshifts to properly catch all mergers; see end of 
\S\ref{sec:mergecrit} for more).
Our redshift range of $z=0.6-5$ allows us to examine subhalos across $7$~Gyr of 
evolution.

To find the subhalos from the phase space data we first generate a catalog of 
halos using the Friends-of-Friends (FoF) algorithm \citep{DEFW} with a linking 
length of $b = 0.168$ times the mean inter-particle spacing.
This partitions the particles into equivalence classes by linking together all 
particles separated by less than $b$, with a density of roughly 
$\rho>3/(2\pi b^3) \simeq 100$ times the background density.
The longer linking length of $b=0.2$ is often used.
However, this linking length is more susceptible to joining together distinct, 
unbound structures and assigning a halo that transiently passes by another as 
a subhalo.
Thus, we use a more conservative linking length, which for a given halo at our 
mass and redshift regime yields a $\sim 15\%$ lower mass than $b=0.2$.\footnote{
Many Millennium subhalo studies use Spherical Overdensity (SO) halos based on 
an FoF(0.2) catalogue, in part to take out the extra structure joined by the 
larger linking length.
\citet{Net07} compares the FoF(0.2) halo centres and those for the SO halos 
used to define the corresponding subhalo populations.
More generally, \citet{Whi01} compares different mass definitions in detail, 
and \citet{CohWhi08} discusses the relation of FoF(0.2), FoF(0.168), SO(180) 
and Sheth-Tormen \citep[][based on $b=0.2$]{SheTor99b} masses at high redshift.}
We keep all FoF groups with more than $32$ particles, and we refer to these 
groups as ``(host) halos''.
Halo masses quoted below are these FoF masses.

When two halos merge, the smaller halo can retain its identity as a ``subhalo''
inside the larger host halo.
We identify subhalos (and sometimes subhalos within subhalos) using a new 
implementation of the {\sl Subfind\/} algorithm \citep{Spr01}.   
We take subhalos to be gravitationally self-bound aggregations of particles 
bounded by a density saddle point.  
After experimentation with different techniques we find this method gives a 
good match to what would be selected ``by eye'' as subhalos.
We use a spline kernel with $16$ neighbours to estimate the density and keep 
all subhalos with more than $20$ particles.
The subhalo that contains the most mass in the halo is defined as the 
central subhalo, all other subhalos in the same halo are satellites.\footnote{
In most cases, the central subhalo in {\sl Subfind\/} is built around the most 
bound and most dense particle in the group.
However, this is not always the case.}
The central subhalo is also assigned all matter within the halo not assigned 
to the satellite subhalos.
The position of a subhalo is given by the location of its most bound particle, 
and the centre of a host halo is defined by the centre of its central subhalo.
For each subhalo we store a number of additional properties including the 
bound mass, velocity dispersion, peak circular velocity, total potential 
energy, and velocity.

\begin{figure}
\begin{center}
\resizebox{3in}{!}{\includegraphics{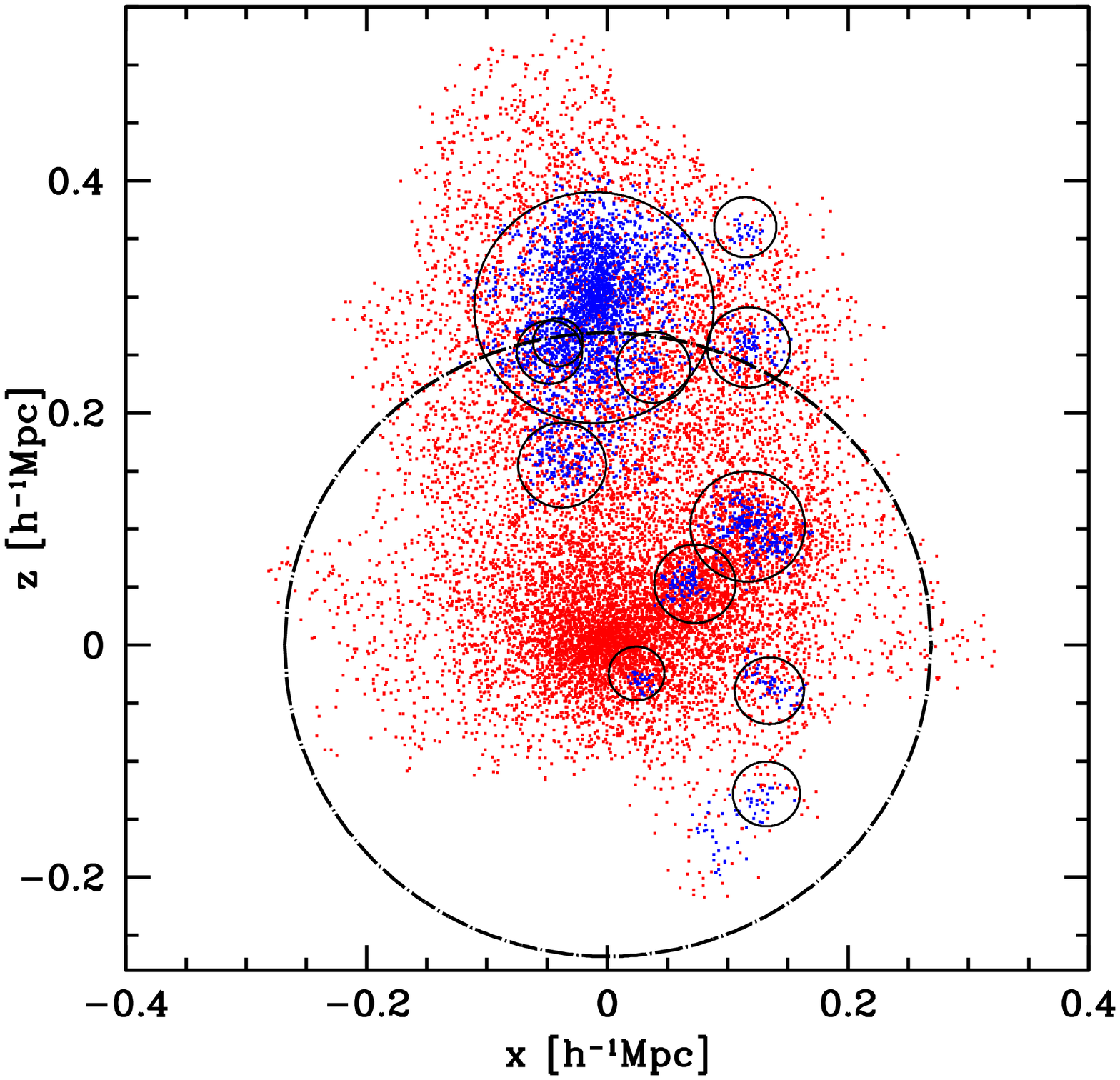}}
\resizebox{3in}{!}{\includegraphics{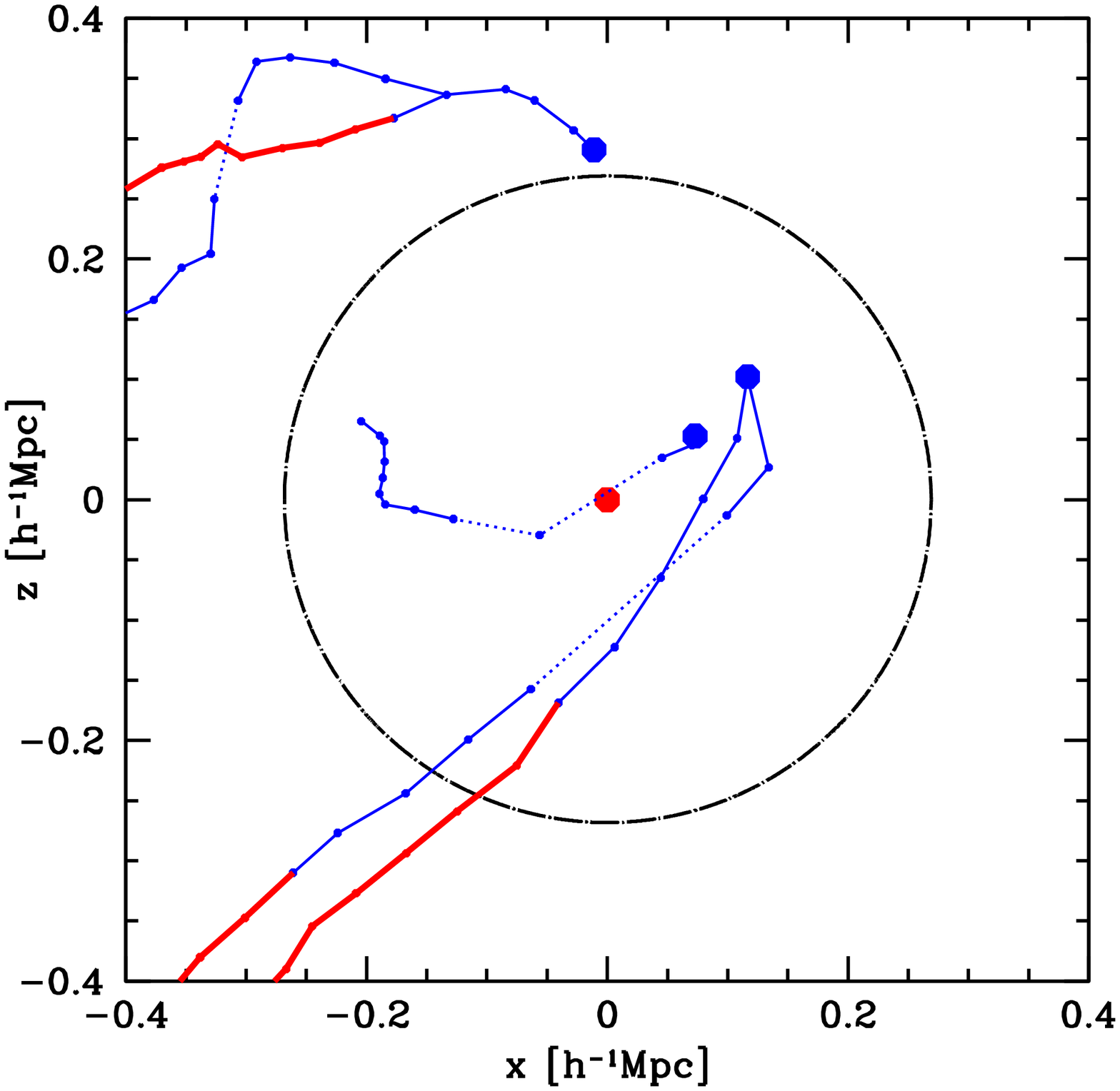}}
\end{center}
\vspace{-0.1in}
\caption{
% iout = 30, igrp = 478068
\textbf{Top}: Projected image of a halo of mass 
$2.2 \times 10^{12} h^{-1}M_\odot$ at $z=2.6$ which hosts $12$ satellite 
subhalos.
Particles assigned to the central subhalo are red, while those assigned to 
satellite subhalos are blue.
Dot-dashed circle shows the halo's virial radius ($r_{200c}$), derived from its
mass assuming a spherical NFW density profile, while the solid circles 
highlight the satellite subhalos and scale in radius with their mass.
The central subhalo has bound mass of $1.8 \times 10^{12} h^{-1}M_\odot$, so 
nearly $20\%$ of the halo's mass lies in satellite subhalos.
\textbf{Bottom}: Tracking histories of massive satellite subhalos in the 
above halo.
Large dots show the positions of satellite subhalo centres at $z=2.6$ for 
subhalos that had a mass $>10^{11} h^{-1}M_\odot$ when they fell into the halo.
Small dots show their positions (relative to that of the halo centre) at 
each output (spaced $50$~Myr) back $800$~Myr.
Thin blue curves show subhalo trajectories when they are satellites while thick 
red curves show when they are centrals (before falling into the halo).
Dotted lines indicate when the parent-child assignment has skipped an output 
(during a fly-by near another subhalo).
} \label{fig:halopicture}
\end{figure}

Figure~\ref{fig:halopicture} (top) shows the projected image of a sample halo 
of mass $2.2 \times 10^{12} h^{-1}M_\odot$, which hosts $12$ satellite subhalos, 
at $z=2.6$ in the $100 h^{-1}$~Mpc simulation.
(At this redshift in our simulations there are approximately $200$ halos per 
($100 h^{-1}$~Mpc)$^3$ above this mass.)
Halos at this mass and redshift regime are dynamically active and often highly
aspherical, with mean axial ratio (ratio of smallest-to-largest semi-major 
axes) of $\sim 0.5$, and recently merged halos have even more discrepant axial 
ratios \citep{All06}.
Figure~\ref{fig:halopicture} also shows how a halo's densest region (figure 
centre), centre of mass, and substructure distribution can all be offset from 
one another.
Because of these asymmetries, satellite subhalos sometimes extend well beyond 
$r_{200c}$, which we will call the halo virial radius.\footnote{
The halo virial radius, $r_{200c}$, i.e. the radius within which the average 
density is $200\times$ the critical density, is calculated from the FoF 
($b=0.168$) mass by first converting to $M_{200c}$ assuming a spherical NFW 
\citep{NFW96} density profile, and then taking 
$M_{200c} = 200 \frac{4\pi}{3} \rho_c r^3_{200c}$.}

\subsection{Subhalo Tracking}

We identify, for each subhalo, a unique ``child'' at a later time, using 
subhalo tracking similar to \citet{Spr05}, \citet{Fal05}, \citet{All05} and 
\citet{Har06}.
We detail our method to illustrate the subtleties which arise and to allow 
comparison with other work.

We track histories over four consecutive simulation outputs at a time because 
nearby subhalos can be difficult to distinguish and can ``disappear'' for a 
few outputs until their orbits separate them again.
For each subhalo with mass $M_1$ at scale factor $a_1$, its child subhalo at a 
later time (scale factor $a_2$ and mass $M_2$) is that which maximizes
\begin{equation}
  \alpha = f\left(M_1,M_2\right)
  		\ln^{-1}\left(\frac{a_2}{a_1}\right)
    \sum_{i\in 2} \phi_{1i}^2
\end{equation}
where
\begin{equation}
f\left(M_1,M_2\right) =
\begin{cases}
	1 - \frac{\left|M_1-M_2\right|}{M_1+M_2} & M_1 < M_2 \\
	1  & M_1 \geq M_2
\end{cases}
\end{equation}
and where $\phi_{1i}$ is the potential of particle $i$ computed using all of 
the particles in subhalo 1, and the sum is over those of the $20$ most bound 
particles in the progenitor that also lie in the candidate child.
We track using only the $20$ most bound particles since our ultimate interest 
is in galaxies, which we expect to reside in the highly bound, central region 
of the subhalo ($20$ is the minimum particle count for our subhalos).
We do not use all the progenitor particles because summing over all of the 
particles in the progenitor that also lie in the child candidate leads to 
instances of parent-child assignment in which the child subhalo does not 
contain the most bound particles of its parent.
Finally, we weight against large mass gains with a mass weighting factor so 
that smaller subhalos passing through larger ones and emerging later on the 
other side are correctly assigned as fly-by's and not mergers.
We find $\sim 95\%$ of subhalos have a child in the next time step, with 
$\sim 4\%$ percent skipping one or more output times and $\sim 1\%$ having no 
identifiable child.

Figure~\ref{fig:halopicture} (bottom) shows the tracking histories of the most
massive subhalos within the halo shown in Fig.~\ref{fig:halopicture} (top), for 
subhalos that had a mass $>10^{11} h^{-1}M_\odot$ when they fell into the halo.
The upper-most subhalo was a separate halo 
($M=1.3 \times 10^{12} h^{-1}M_\odot$) hosting its own massive satellite subhalo.
It then fell into the main halo and then both (now satellite) subhalos merged 
with each other.
The right-most subhalo is an example of two separate halos falling into the 
main halo, becoming satellites, and subsequently merging with each other.
Finally, the track through the centre shows a single subhalo falling towards
the central subhalo.
Instead of merging with the central, it passes through as a fly-by.

All of these tracks show instances where the parent-child assignment algorithm 
has skipped an output (dotted lines) as one subhalo passes through another and 
re-emerges on the other side.
Note also that, while the dot-dashed circle shows the halo virial radius,
$r_{200c}$ (at the last output), based on its mass assuming a spherical NFW 
density profile, the locations of the transitions of central subhalos to 
satellites during infall show that the spherical virial radius is only a rough 
approximation.

\begin{figure}
\begin{center}
\resizebox{3in}{!}{\includegraphics{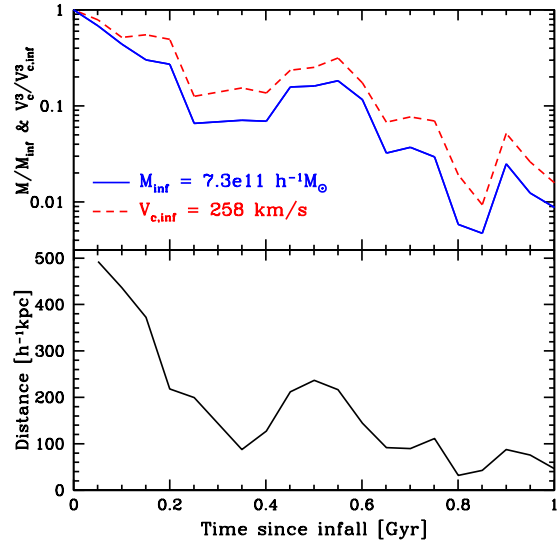}}
\end{center}
\vspace{-0.1in}
\caption{
% iout = 28, igrp = 432087
\textbf{Top}: Evolution of mass (solid curve) and circular velocity (dashed 
curve) as a function of time since infall for a satellite that fell into a halo 
of mass $8 \times 10^{12} h^{-1}M_\odot$ at $z=3.7$.
\textbf{Bottom}: Radial distance of satellite from centre of host halo.
Mass and circular velocity exhibit correlations with radial distance,  
such as mass gain as the satellite recedes from the centre of its host halo.
This satellite experienced no major merger activity throughout its history.
} \label{fig:satevol}
\end{figure}

As a halo falls into a larger host halo and becomes a satellite subhalo, mass 
loss from tidal stripping can be extreme ($90\%$ or more) as the satellite 
subhalo orbits towards the centre of its host halo.
Figure~\ref{fig:satevol} shows the evolution from infall of a long-lived 
satellite with infall mass $M_{\rm inf} = 6 \times 10^{11} h^{-1}M_\odot$ and 
maximum circular velocity at infall of $V_{\rm c,inf} = 258$~km/s.
Mass and maximum circular velocity ($V_{\rm c,max}^3$) are stripped by 
$\sim 90\%$ by the time the satellite first reaches pericentre.
After pericentric passage, a satellite can also gain mass and circular velocity
as it moves away from the dense centre of its host halo.
This is shown in Fig.~\ref{fig:satevol} where mass and circular velocity 
strongly correlate with radial distance throughout the orbit.
This is partially an effect of our subhalo finder: the density contrast of a 
subhalo, which defines its physical extent and hence its mass, drops as it moves
to a region of higher background density.\footnote{
A fly-by, i.e. when a subhalo passes through and is temporarily 
indistinguishable from a larger subhalo, is an extreme case of this.}
However, this is also driven by a physical effect: as a satellite approaches 
the dense centre of its host halo, it will be compressed by tidal shocks in 
response to the rapidly increasing potential 
\citep[see Fig.~12 of][]{DieKuhMad07,GneOst97,DekDevHet03}.

Since a central subhalo is defined as the most massive subhalo, a satellite 
subhalo can become a central (and vice versa), which we refer to as a 
``switch''.\footnote{
In more detail, a switch occurs when the density peak of a satellite (above the 
background) contains more mass than is within the central subhalo's radius at 
the position of the satellite}.
We find that $\sim 4\%$ of all centrals at any output become satellites within 
the same halo in the next output, and of these a third immediately return to 
being centrals in the following output.
Switches are twice as common during a merger event, either when two satellites 
merge, or when a satellite merges with the central, forming a less dense object 
and allowing another satellite to bind more mass to itself.
When a satellite switches to a central and back to a satellite, the original
satellite can be mistakenly assigned as a direct parent of the final satellite 
(and thus the central is assigned no parent) since our child assignment weights 
against large mass gains (which occurs when a satellite becomes a central).
We fix these distinct cases by hand.

These switches highlight the fact that the distinction between a central and 
satellite subhalo is often not clear-cut: at this redshift and mass regime  
massive halos undergo rapid merger activity and thus often are highly disturbed 
and aspherical, with no well-defined single peak that represents the center of 
the halo profile.

\subsection{Subhalo Mass Assignment}

Since we use subhalos as proxies for galaxies, we track subhalo mass that 
is expected to correlate with galaxy stellar mass.
Galaxies form at halo centres as baryons cool and adiabatically contract 
toward the minimum of the halo's potential well, which leads to a correlation 
between halo mass and galaxy stellar mass \citep{WhiRee78,Blu86,Dub94,
MoMaoWhi98}.
When a halo falls into a larger halo and becomes a satellite subhalo, its 
outskirts are severely stripped as discussed above, but its galaxy's stellar 
mass would be little influenced as the galactic radius is typically 
$\sim 10\%$ that of the subhalo radius.
This motivates assigning to subhalos their mass at infall, $M_{\rm inf}$, 
which is expected to correlate with galactic stellar mass throughout the 
subhalo's lifetime.
The subhalo infall mass function has been successful at reproducing the 
observed galaxy luminosity function and clustering at low redshifts
\citep{ValOst06,WanKauDeL06,YanMovdB08}.
Maximum circular velocity at infall, $V_{\rm c,inf}$, has also been successfully
matched to some observations \citep{ConWecKra06,BerBulBar06}.
In Appendix A we show the relation between $M_{\rm inf}$ and $V_{\rm c,inf}$ and 
its redshift evolution.

Our prescription for assigning $M_{\rm inf}$ to subhalos is as follows.
When a halo falls into another and its central subhalo becomes a satellite 
subhalo, the satellite is assigned $M_{\rm inf}$ as the subhalo mass of its 
(central) parent.\footnote{
There is some dependence on output time spacing in this definition: since a 
central subhalo typically continues to gain mass before it falls into a 
larger halo, shorter time steps (which catch it closer to infall) lead to 
a higher subsequent satellite $M_{\rm inf}$.
Doubling the output spacing leads to satellites with $\sim10\%$ lower infall 
mass.}
If a satellite merges with another satellite, the resultant child subhalo is 
assigned the sum of its parents' $M_{\rm inf}$.
Since the central subhalo contains the densest region of a halo, inter-halo gas 
is expected to accrete onto it, so we define $M_{\rm inf}$ for a central subhalo 
as its current self-bound subhalo mass, which is typically $\sim 90\%$ of its 
host halo's mass.\footnote{
Though because of finite resolution, we find a weak systematic drop in 
$M_{\rm cen}/M_{\rm halo}$ with halo mass, varying from $93\%$ to $87\%$ 
for $10^{11}$ to $10^{14} h^{-1}M_{\odot}$ in the higher resolution simulation.}
However, since a central subhalo can switch to being a satellite, while a 
satellite switches to being a central (all within a single host halo), we 
require an additional rule because using the above simple assignment of 
$M_{\rm inf}$ to centrals would lead to a central and satellite in a halo each 
having the halo's current bound mass.
Thus, we assign a central to have $M_{\rm inf}$ as its current self-bound 
subhalo mass only if it was the central in the same halo in the previous output.
Thus, if a satellite switches to a central and remains the central for multiple 
outputs, it has robustly established itself as the central subhalo, so it is 
assigned its current self-bound mass.
However, if a central was a satellite (or a central in another smaller halo) 
in the previous output, it is assigned the sum of its parents' $M_{\rm inf}$,
with the additional requirement that its $M_{\rm inf}$ cannot exceed its 
current self-bound mass.

A small fraction ($\sim 4\%$) of satellites composed of at least $50$ 
particles are not easily identifiable with any progenitor subhalos and thus 
cannot be tracked to infall.
On inspection, we find that these ``orphaned'' satellites are loosely 
self-bound portions of a central subhalo, remnants from a collision between a 
satellite and its central subhalo that soon re-merge with the central.
Given their origins and fates, these orphans are not expected to host galaxies 
and are ignored.

Although we track subhalos down to $20$ particles, we impose a much larger 
minimum infall mass to our sample to avoid selecting subhalos that artificially 
dissolve and merge with the central too early.
This requires sufficient resolution of the radial density profile of a 
satellite subhalo at infall: if the satellite's core is smaller than a few 
times the force softening length, its profile will be artificially shallow and 
it will be stripped and disrupted prematurely (see Appendices for details).
For calibration, we use the regime of overlap in mass between our two 
simulations of different mass resolution, requiring consistent subhalo mass 
functions, halo occupation distributions, and merger statistics for a minimum 
infall mass.
For example, going too low in mass for the larger simulation resulted in more 
mergers and fewer satellites than for the same mass range in the smaller 
simulation.
In the larger simulation, our consistency requirements led us to impose
$M_{\rm inf} > 10^{12} h^{-1}M_\odot$.
For a fixed number of particles per subhalo, this scales down to 
$M_{\rm inf} > 10^{11} h^{-1}M_\odot$ in the smaller, higher resolution 
simulation.
Note that halos of mass $10^{11}~(10^{12}) h^{-1}M_\odot$ cross below $M_*$, the 
characteristic mass of collapse, at $z=1.5~(z=0.8)$, so we probe massive 
subhalos across most of our redshift range.

At $z=2.6$, there are $\sim 16000$ subhalos with 
$M_{\rm inf} > 10^{11} h^{-1}M_\odot$ in our $100 h^{-1}$~Mpc simulation and 
$9400$ subhalos with $M_{\rm inf} > 10^{12} h^{-1}M_\odot$ in our 
$250 h^{-1}$~Mpc simulation.
At $z=1$, there are $\sim 29400$ ($2500$) subhalos with 
$M_{\rm inf} > 10^{11}~(10^{12}) h^{-1}M_\odot$ in our $100 h^{-1}$~Mpc simulation.

\subsection{A Note on Stellar Mass and Gas Content of Subhalo Galaxies}

A galaxy's stellar mass is expected to be a non-linear, redshift-dependent
function of its subhalo mass.
An approximate relation based on abundance matching is given in 
\citet{ConWec08}.
At $z=1$, they find that subhalos of infall mass
$10^{11}~(10^{12}) h^{-1}M_\odot$ host galaxies of stellar mass 
$\sim 10^{9}~(10^{10.5}) M_\odot$.
At $z=2.5$, subhalos at the above masses are expected to host lower mass 
galaxies, though quantitative relations at this redshift are less certain, and 
our subhalo and halo finders differ in detail from theirs.
Our sample of $M_{\rm inf} > 10^{12} h^{-1}M_\odot$ subhalos approximately 
corresponds to $L \gtrsim L_*$ galaxies at the redshifts we examine.

Although our simulations do not track the baryonic content of subhalos, most 
massive galaxies are gas-rich at high redshift.
That is, at $z \approx 1$, $70\%$ to $90\%$ of $L \sim L_*$ galaxies are 
observed to be blue \citep{Coo07,Ger07}, possessing enough gas to be actively 
star forming.
The fraction of gas-rich galaxies at higher redshift is more poorly constrained 
but is thought to be higher \citep{HopCoxKer08}.
Thus, we anticipate that most, if not all, mergers we track have the capacity 
to drive galaxy activity such as starbursts and quasars.

\subsection{Satellite Fraction} \label{sec:satfrac}

\begin{figure}
\begin{center}
\resizebox{3in}{!}{\includegraphics{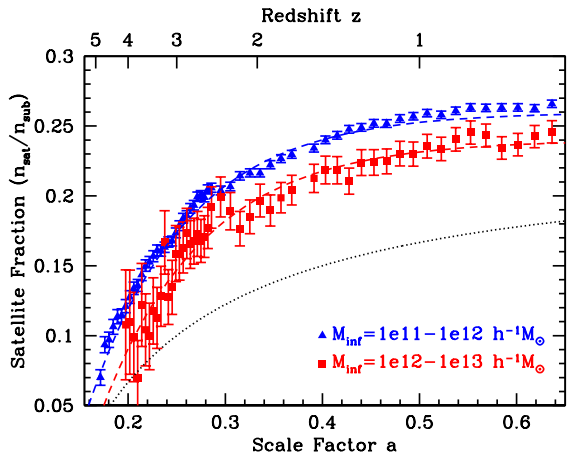}}
\end{center}
\vspace{-0.1in}
\caption{
Evolution of the fraction of all subhalos that are satellites for 
$M_{\rm inf} = 10^{11}-10^{12} h^{-1}M_\odot$ (triangles) and 
$M_{\rm inf} = 10^{12}-10^{13} h^{-1}M_\odot$ (squares).
Dashed lines show fits to Eq.~\ref{eq:satfracA}.
Dotted line shows fit to Eq.~\ref{eq:satfracB} (that of \citet{ConWec08}).
} \label{fig:satfrac}
\end{figure}

To frame our ensuing discussion of mergers, and the relative importance of
satellite and central subhalos, Fig.~\ref{fig:satfrac} shows the evolution of 
the satellite fraction, $n_{\rm satellite}/n_{\rm subhalo}$, for subhalos of a 
fixed mass\footnote{
We show the satellite fraction for two mass bins, but since the mass function 
falls exponentially at these masses and redshifts, almost all objects are 
at the low end of the mass bin.
Using instead a minimum mass cut changes our results by only a few percent.}, 
which grows monotonically with time from $z=5-0.6$, as
\begin{equation} \label{eq:satfracA}
\frac{n_{\rm satellite}}{n_{\rm subhalo}} = C-e^{-\frac{\beta}{1+z}}
\end{equation}
with $\beta = 9.7$ valid across all subhalo masses and $C = 0.26~(0.24)$ for 
lower (higher) mass.
The satellite fraction decreases with increasing subhalo mass since more 
massive subhalos are more likely to be centrals, and for a fixed subhalo mass 
the satellite fraction increases with time as the number of high mass halos 
hosting massive satellites increases.
The increase in the satellite fraction is slowed at late times because the 
number of massive satellites in halos of a fixed mass decreases with time as 
the satellites coalesce with the central subhalo (see \S\ref{sec:hodevol}).

Note that Eq.~\ref{eq:satfracA} predicts a much higher satellite fraction than 
that of \citet{ConWec08}, who found
\begin{equation} \label{eq:satfracB}
\frac{n_{\rm satellite}}{n_{\rm subhalo}} = 0.2-\frac{0.1}{3}z .
\end{equation}
It is unlikely that the difference is driven by numerical effects, since both 
their and our simulations are of similar mass resolution and volume, and both 
analyses are based on similar subhalo infall mass cuts (we see similar satellite
fractions selecting instead on fixed $V_{\rm c,inf}$).
One likely factor is that, as noted above, our smaller output spacing yields 
higher infall mass for satellite subhalos, since we catch halos closer to 
infall when their mass is higher.
In addition, they use a different halo and subhalo finding algorithm
\citep{KlyGotKra99,KraBerWec04}, and their higher $\sigma_8$ ($0.9$ rather 
than our $0.8$) would give a lower merger rate and thus fewer satellites (one 
expects satellite survival timescales not to change).
If normalized to the same satellite fraction at a given epoch, the redshift 
evolution of Eqs.~\ref{eq:satfracA} and \ref{eq:satfracB} are in rough 
agreement.

%% MERGER RATES %%%%%%%%%%%%%%%%%%%%%%%%%%%%%%%%%%%%%%%%%%%%%%%%%%%%%%%%%%%%%%%%
\section{Merger Criteria and Rates} \label{sec:rates}

\subsection{Merger Criteria}
\label{sec:mergecrit}

For two parents with $M_{\rm inf,2} \leq M_{\rm inf,1}$ sharing the same child 
subhalo at the next output they appear, a child is flagged as a (major) merger 
if $\frac{M_{\rm inf,2}}{M_{\rm inf,1}} > \frac{1}{3}$.
If a child has more than two parents, we count multiple mergers if any other 
parents also exceed the above mass ratio with respect to the most massive 
parent.

Our mass ratio represents a trade-off between strong mergers (to maximize 
signal) and frequent merging (for statistical power).
Galaxy mergers with stellar mass ratios closer than 3:1 are expected to drive 
interesting activity, e.g. quasars and starbursts, as mentioned above.\footnote{
Since galaxy stellar mass is a non-linear function of subhalo mass, a subhalo 
mass ratio of 3:1 may correspond to a galaxy merger of a more or less 
discrepant mass ratio.
However, since the $M_{\rm stellar}-M_{\rm subhalo}$ relation is expected to 
peak for $M_{\rm subhalo} \sim 10^{12} h^{-1}M_\odot$ \citep{ConWec08}, we do 
not expect this effect to strongly bias our results.}

For generality, the distribution of merger (infall) mass ratios, $R$:1, for 
both halos and subhalos in the mass and redshift regimes we consider can be
approximated by
\begin{equation} \label{eq:mergeratio}
f(R) \propto R^{-1.1}
\end{equation}
in reasonable agreement with the $R^{-1.2}$ distribution of galaxy mass ratios 
at $z<0.5$ found in hydrodynamic simulations by \citet{Mal06}.\footnote{
The distribution of merger mass ratios also agrees well with the fit for halos 
at $z=0$ provided by \citet{WetSchHol08}.}
Thus, the counts/rates of mergers with (infall) mass ratio closer than $R$:1 can
be approximately scaled from our results through the relation
\begin{equation} 
N(<R) = 8.6  \, (R^{0.1}-1)N(<3) .
\end{equation}

Sometimes merger criteria are based on instantaneous subhalo mass gain 
\citet[e.g.,][]{ThaScaCou06}.
However, as exemplified in Fig.~\ref{fig:satevol}, subhalos can gain 
significant mass without coalescence.
We find that most cases of significant subhalo mass gain are not a two-body 
coalescence, and so we do not use this to select mergers (see 
\citet{WetCohWhi09b} for more detail, also related results in \citet{Mau07}).

Sufficient output time resolution is necessary to properly resolve the merger 
population.
Unless otherwise stated, we use the shortest simulation output spacing to 
define the subhalo merger time interval, corresponding to $50$~Myr 
($\sim 100$~Myr) for $M_{\rm inf} > 10^{11}~(10^{12}) h^{-1}M_\odot$ 
at $z>2.5$, and  $\sim 200$~Myr for all masses at $z<1.6$.
By tracking parent/child subhalo assignments across multiple output spacings, 
we found that longer output time spacings result in a lower subhalo merger 
rate, arising from a combination of effects.
As stated above, shorter time steps catch halos closer to their infall, 
resulting in satellites with higher $M_{\rm inf}$, and hence more major 
satellite-central mergers.
In addition, satellite-satellite mergers can be missed if their child merges 
into the central subhalo before the next time step.
However, smaller output spacings are also more susceptible to switches, which 
can yield satellites with higher mass.

For halos, very short time steps can artificially enhance the merger rate, 
catching transient behavior as halos intersect, pass through each other, and 
re-merge.
While we use FoF(0.168) instead of FoF(0.2) halos to help minimize artificial 
bridging effects, for a $50$~Myr time spacing halo re-mergers still constitute 
a significant fraction of halo mergers.
Thus we calculate all our halo mergers using $>100$~Myr output spacings, for 
which halo re-mergers constitute only a few percent of all mergers.

Given our simulation volumes we are unable to probe statistically significant 
subhalo merger counts in halos more massive than 
$4\times 10^{13} h^{-1}M_\odot$ at $z>2.5$ and $M \sim 10^{14} h^{-1}M_\odot$ 
at $z<1.6$.
However, the majority of recently merged subhalos are centrals (discussed 
below), and selecting on higher mass host halos for a fixed subhalo mass 
restricts increasingly to recently merged satellites (of which there are 
comparatively fewer).
Furthermore, nearly all subhalo mergers at a given mass occur in halos at most 
$5-10\times$ times more massive.

\subsection{Fits to Simulation}

\begin{figure*}
\begin{center}
\resizebox{3in}{!}{\includegraphics{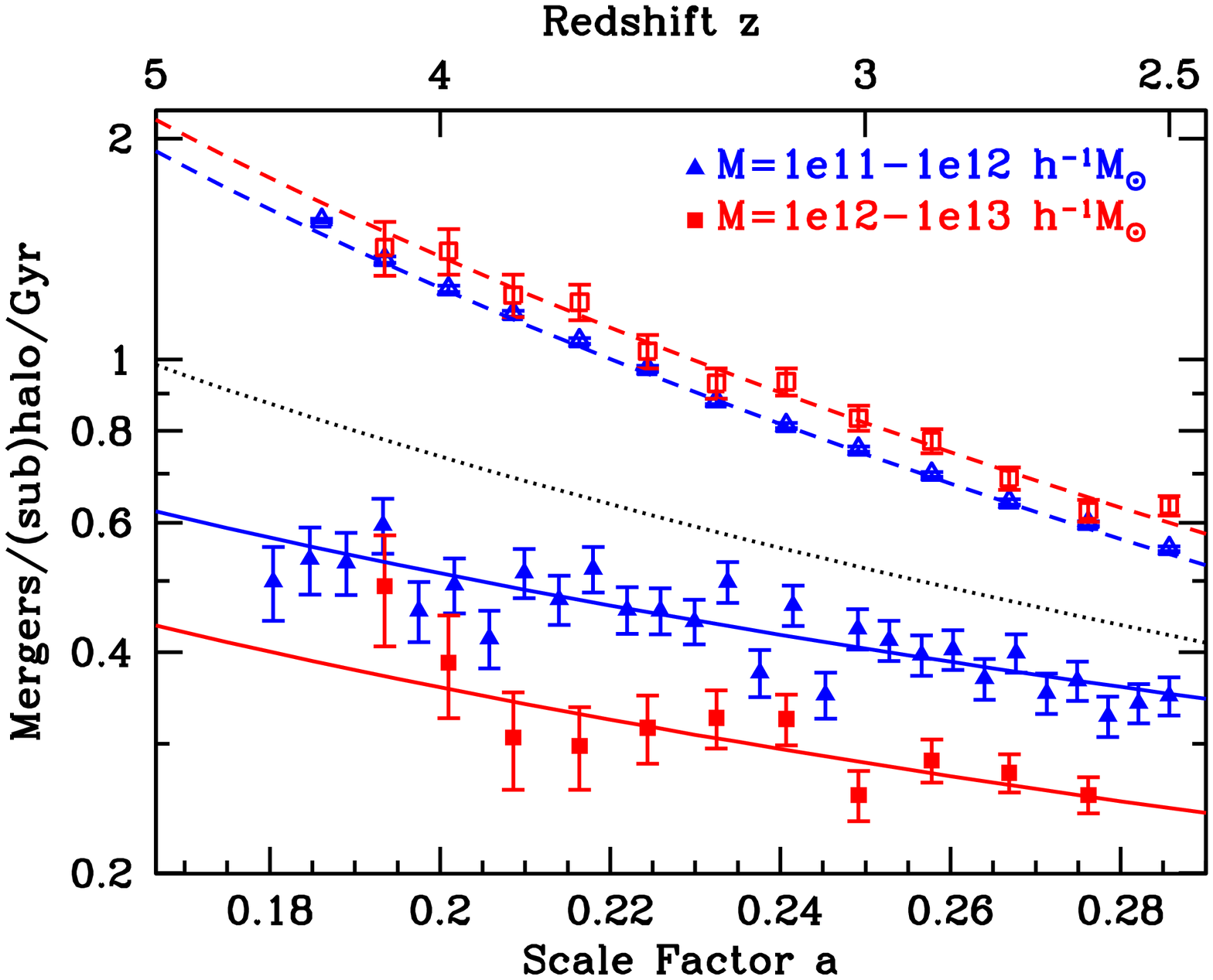}}
\resizebox{3in}{!}{\includegraphics{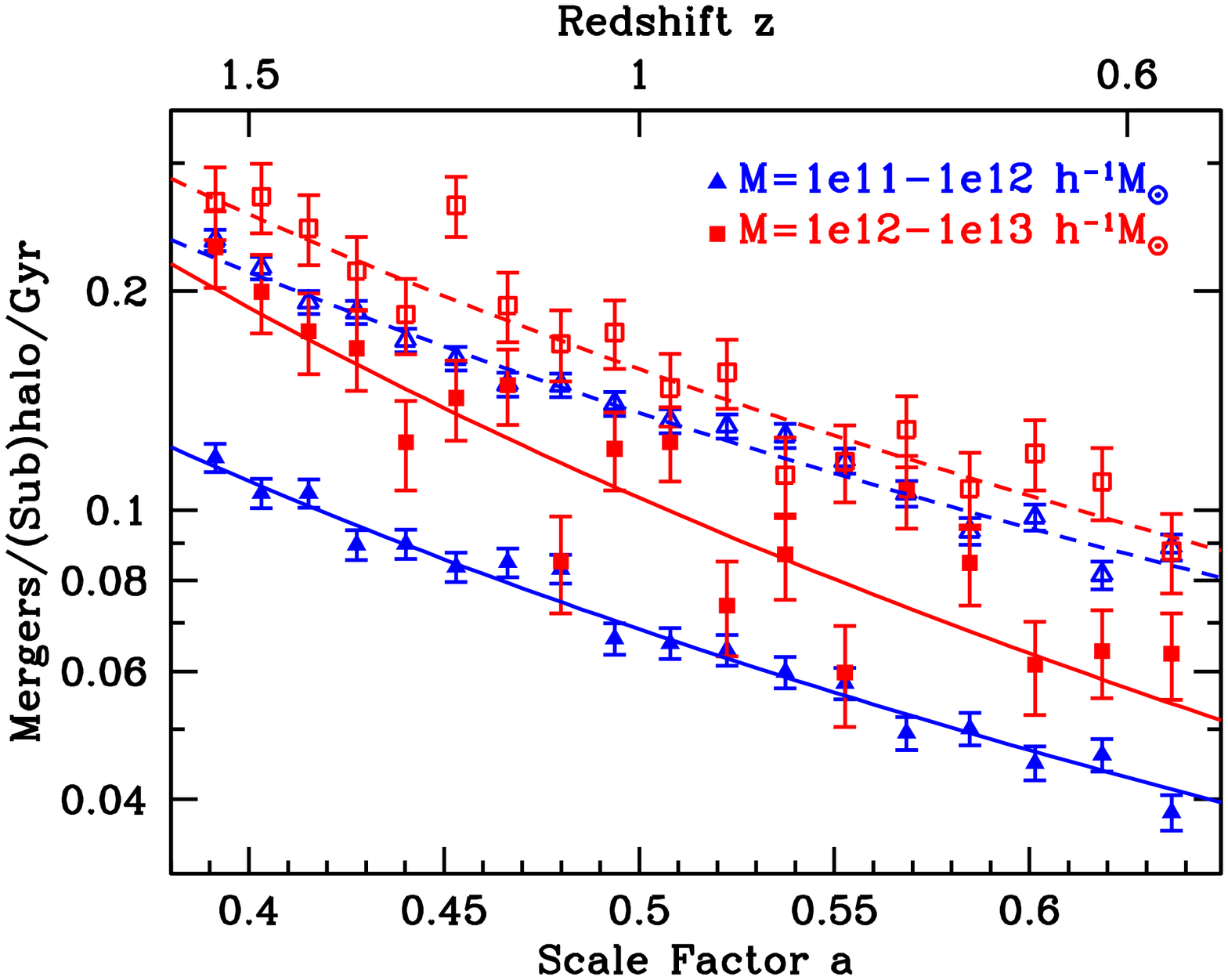}}
\end{center}
\vspace{-0.1in}
\caption{
\textbf{Left}: Merger rate per object of subhalos (solid points) and halos 
(open points) as a function of the scale factor, for 
$M_{\rm inf} = 10^{11}-10^{12} h^{-1}M_\odot$ (triangles) and 
$M_{\rm inf} = 10^{12}-10^{13} h^{-1}M_\odot$ (squares) at $z=5-2.5$.
Solid (dashed) curves show fits of subhalo (halo) merger rates to 
Eq.~\ref{eq:mergerate}, with fit parameters given in Table~\ref{tab:mergefits}.
\textbf{Right}: Same, but at $z=1.6-0.6$.
Also shown at left is the fit to the halo merger rate for halos with a central 
subhalo of $M_{\rm inf} = 10^{11}-10^{12} h^{-1}M_\odot$ (dotted).
} \label{fig:mergerates}
\end{figure*}

\begin{table*}
\begin{center}
\begin{tabular}{|l|c|c|c|c|c|}
\hline \multicolumn{2}{|c|}{Redshift} & \multicolumn{2}{|c|}{$z=5-2.5$} & 
\multicolumn{2}{|c|}{$z=1.6-0.6$} \\
\hline \multicolumn{2}{|l|}{Mass [$h^{-1}M_\odot$]} & $10^{11}-10^{12}$ & 
$10^{12}-10^{13}$ & $10^{11}-10^{12}$ & $10^{12}-10^{13}$ \\
\hline \multirow{2}{*}{Halos} & A & 0.029 & 0.032 & 0.034 & 0.034 \\ 
	& $\alpha$ & 2.3 & 2.3 & 2.0 & 2.2 \\
\hline \multirow{2}{*}{Subhalos} & A & 0.093 & 0.065 & 0.016 & 0.016 \\ 
	& $\alpha$ & 1.1 & 1.1 & 2.1 & 2.6 \\
\hline
\end{tabular}
\end{center}
\caption{
The amplitude, $A$, and power law index, $\alpha$, for halo and subhalo merger
rates fit to Eq.~\ref{eq:mergerate}.
} \label{tab:mergefits}
\end{table*}

Figure~\ref{fig:mergerates} shows the evolution of the merger rate per 
object---defined as the number of mergers per time per object---for both 
subhalos and halos of the same (infall) mass.\footnote{
Merger fraction errors are calculated using binomial statistics.
Given $M$ mergers out of $N$ objects, the mostly likely fraction is $f = M/N$
with variance $\sigma_f^2 = \frac{M(N-M)+1+N}{(N+2)^2(N+3)}$.}
We fit the merger rate per object as 
\begin{equation} \label{eq:mergerate}
	\frac{n_{\rm merge}}{n_{\rm obj} dt} = A(1+z)^{\alpha}
\end{equation}
where $n_{\rm merge}$ is the number of (sub)halos whose parents match our mass 
ratio selection, $n_{\rm obj}$ is the total number of (sub)halos within the 
same mass range at the same output, and $dt$ it the time interval between
consecutive outputs.
The best-fit values in each redshift regime are shown in 
Table~\ref{tab:mergefits}.
Note that the merger rate we examine is the number of mergers per time
\textit{per object}, different from another common definition, the number of 
mergers per time \textit{per volume}.\footnote{
The former is simply the latter divided by the number density of objects of the 
same mass, but the merger rate per volume has qualitatively different behaviour 
because the (comoving) number density of objects at a fixed mass increases with 
time in a redshift-dependent manner.
Specifically, at high redshift where the mass function rapidly increases, the 
merger rate per volume for both halos and subhalos \textit{increases} with 
time, reaching a peak at $z \sim 2.5$.
Below this redshift, it decreases with time in a power law manner as in 
Fig.~\ref{fig:mergerates}, though with a shallower slope of 
$\alpha \approx 1.5$.}

The relation of subhalo mergers to halo mergers is nontrivial.
Although subhalo mergers are the eventual result of halo mergers, the former 
are governed by dynamics within a halo and the latter by large-scale 
gravitational fields.
For halos, both the slope and the amplitude of the merger rate exhibit little 
dependence on mass or redshift.
Similarly, e.g. \citet{FakMa08} found $\alpha = 2-2.3$ for all halo masses at 
$z<6$, and weak halo mass dependence of the amplitude.
In contrast, the subhalo merger rate amplitude has strong dependence on mass 
and its slope depends strongly on redshift.
Relative to the halo merger rate, the subhalo merger rate is lower in amplitude 
than that of halos of the same (infall) mass, and, most notably at $z>2.5$, the 
rate of subhalo mergers falls off significantly more slowly that that of halos.
This is consistent with earlier work: \citet{DeL04} found a higher merger 
fraction for halos than subhalos, and \citet{GuoWhi08} found strong mass 
dependence of the galaxy merger rate amplitude and that the slope becomes much 
shallower at $z>2$ \citep[see also][]{Mat08}.
(Note that as our halos are FoF(0.168) halos and so our merger rates can differ 
from those for FoF(0.2) halos; merging occurs sooner for a finder with a larger 
linking length.)

We now focus on the relation between halo and subhalo merger rates to 
understand these trends with time.

\subsection{Subhalo vs. Halo Merger Rates}

A simple analytic argument based on dynamical infall time, i.e., subhalo 
mergers are simply a delayed version halo mergers, leads one to expect that 
subhalo merger rates simply track those of halos: they have the same time 
evolution, with the subhalo merger rate having a higher amplitude.
In this argument, when two halos merge, the new satellite galaxy collides with 
the other central galaxy within a dynamical friction timescale, approximated by 
\begin{equation} \label{eq:dynfric}
	t_{\rm merge} \approx C_o\frac{M_{\rm halo}/M_{\rm sat}}{\ln(1+M_{\rm halo}/M_{\rm sat})}t_{\rm dyn}
\end{equation}
where $t_{\rm dyn} = 0.1 t_{\rm Hubble}$, $t_{\rm Hubble} = \frac{1}{H(z)}$, and 
$C_o \approx 1$ accounts for the ensemble averaged satellite orbital parameters 
\citep{ConHoWhi07,BinTre,BoyMaQua08,Jia08}.
Thus, for a fixed mass ratio, letting 
$m_o = M_{\rm halo}/M_{\rm sat}/\ln(1+M_{\rm halo}/M_{\rm sat})$,
\begin{equation}
	t_{\rm sat,merge} \approx 0.1 \, C_o m_o t_{\rm Hubble} \approx 0.1 \, C_o m_o t .
\end{equation}

The evolution of the halo merger rate per object during matter-domination 
(valid at the high redshifts we examine), where $a \propto t^{2/3}$, is 
approximately
\begin{equation}
	\frac{n_{\rm merge}}{n_{\rm halo} dt} = A(1+z)^{\alpha} = \left(\frac{t}{t_*}\right)^{-\frac{2}{3}\alpha} ,
\end{equation}
with $t_*$ some proportionality constant.
Assuming all halo mergers lead to satellite-central subhalo mergers on a 
dynamical friction timescale, the subhalo merger rate per object would evolve 
with time as 
\begin{align}
\frac{n_{\rm merge}}{n_{\rm subhalo}dt} &= \left(\frac{t-t_{\rm sat,merge}}{t_*}\right)^{-\frac{2}{3}\alpha} \\
&= {(1-0.1 \, C_om_o)^{-\frac{2}{3} \alpha}}\left(\frac{t}{t_*}\right)^{-\frac{2}{3} \alpha} 
\end{align}
and so the subhalo merger rate would simply track that of halos of the same 
mass, but with higher amplitude.

\subsection{Resolving the Discrepancy}

As Fig.~\ref{fig:mergerates} shows, however, this tracking does not occur, 
particularly at high redshift where the slope of the subhalo merger rates is 
much shallower than that of halos.
One reason for this is that we compute the merger rate per object, in which we
divide by the number of objects at the given mass, $n_{\rm obj}(m)$.
Since halo masses are added instantaneously during halo mergers, a recently 
merged halo will instantly jump to a higher mass regime (with smaller
$n_{\rm obj}$ in Eq.~\ref{eq:mergerate}).
In contrast, the central subhalo of the resultant halo will remain at a smaller 
mass for some time until its mass grows from stripping of the new satellite 
subhalo (i.e., recently merged halos have lower than average 
$M_{\rm inf,cen}/M_{\rm halo}$), so the central's $n_{\rm obj}$ is higher.

In Fig.~\ref{fig:mergerates} (left) the dotted black curve fits the halo merger 
rate for halos selected with the same mass cut on their central subhalo infall 
mass.
The amplitude is significantly smaller, and the slope is also shallower 
($\alpha = 1.6$) than for halos selected on full halo mass, showing that the 
above effect is stronger at earlier times.
By the argument above, the ratio of amplitudes of the merger rate per object of 
halos to subhalos is $n_{\rm halo}/n_{\rm cen}$.
A fixed (sub)halo mass cut probes lower $M/M_*(t)$ at later times, and the mass 
function drops exponentially with increasing $M/M_*(t)$ at these masses.
Thus, a fixed $M_{\rm inf,cen}/M_{\rm halo}$ after a halo merger means
$n_{\rm halo}/n_{\rm cen}$ becomes closer to unity at later times, leading to 
the shallower slope.

The measured subhalo mergers have an even lower amplitude and shallower slope 
than the dotted line in Fig.~\ref{fig:mergerates}, driven by two additional
effects.
First, a halo major merger might not lead to a subhalo major merger since the 
satellite-central merger mass ratio can be smaller than the mass ratios of 
their source halos.
This is because $M_{\rm inf}$ naturally grows for a central but only grows for 
a satellite if it has a merger before coalescing with the central 
\citep[see][for a detailed analysis of this effect in terms of assigning 
baryons to subhalos]{WanKau08}.
Since halos grow in mass more quickly at higher redshift, this effect is 
stronger at earlier times, further flattening the subhalo merger rate slope.
Second, there is a significant contribution of recently merged satellites to 
the merger population.
We find that satellites are twice as likely to have had a recent merger as 
centrals of the same mass, regardless of mass cut and redshift (see
\S\ref{sec:types} for more detail).
This enhances the merger rate, with a stronger enhancement at later times 
since the satellite fraction grows with time as in Fig.~\ref{fig:satfrac}.

At lower redshift, Fig.~\ref{fig:mergerates} (right) shows that the amplitude 
of the subhalo merger rate remains lower than that of halos, but the slopes 
become similar, indicating the effects examined above become less 
time-dependent.
Since the masses we probe are crossing $M_*(t)$ as these redshifts, the reduced 
amplitude from subhalo vs. halo mass cut becomes less sensitive to time.
Similarly, our mass range crossing $M_*(t)$ means that halo mass growth slows, 
so the fraction of halo major mergers that leads to subhalo major mergers 
remains roughly constant with time.
Finally, as shown in Fig.~\ref{fig:satfrac}, the satellite fraction growth
asymptotes at lower redshift, which means that the enhancement from recently 
merged satellites remains roughly constant.

%% MERGER TYPES %%%%%%%%%%%%%%%%%%%%%%%%%%%%%%%%%%%%%%%%%%%%%%%%%%%%%%%%%%%%%%%%
\section{Satellite vs. Central Mergers} \label{sec:types}

\begin{figure*}
\begin{center}
\resizebox{3in}{!}{\includegraphics{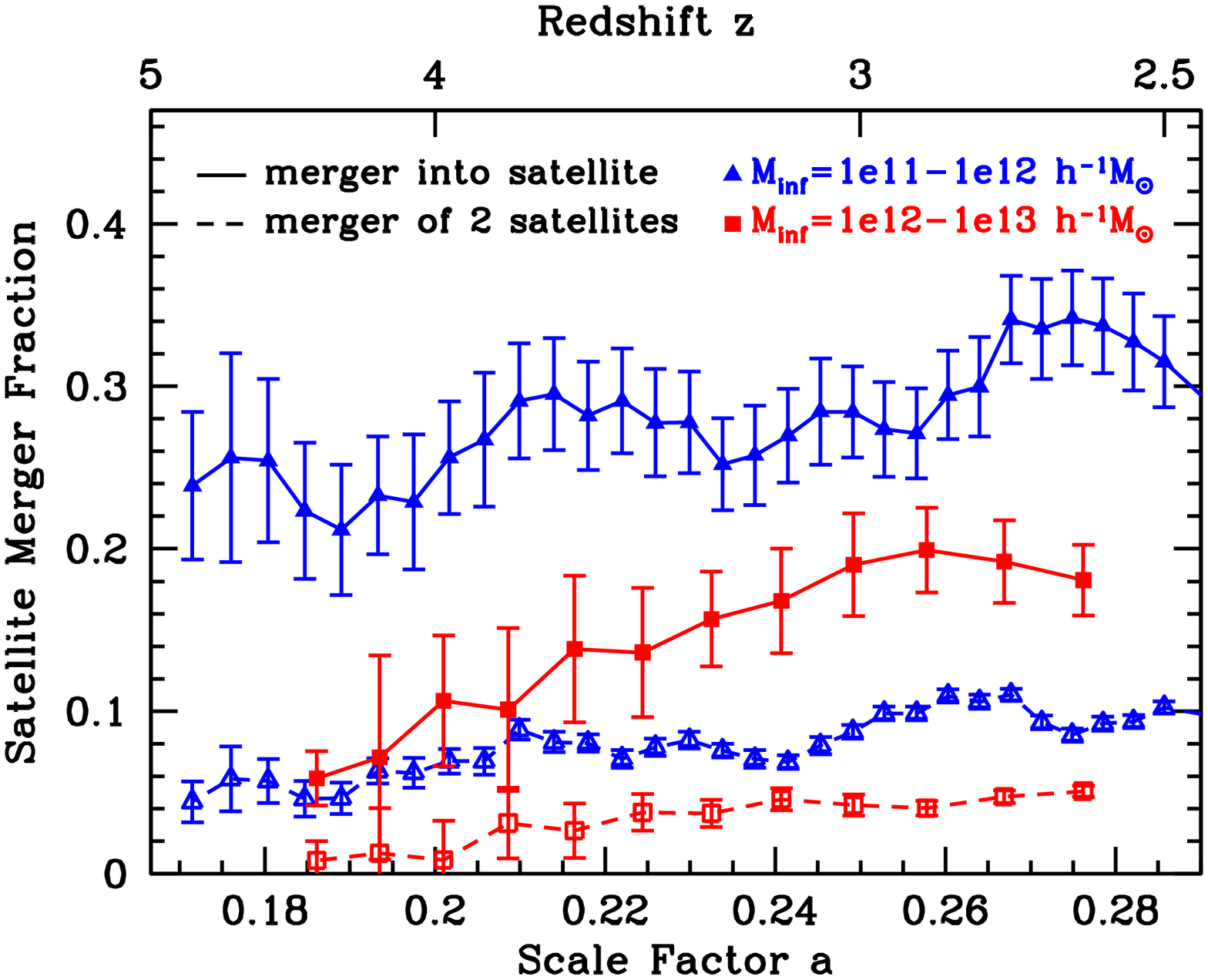}}
\resizebox{3in}{!}{\includegraphics{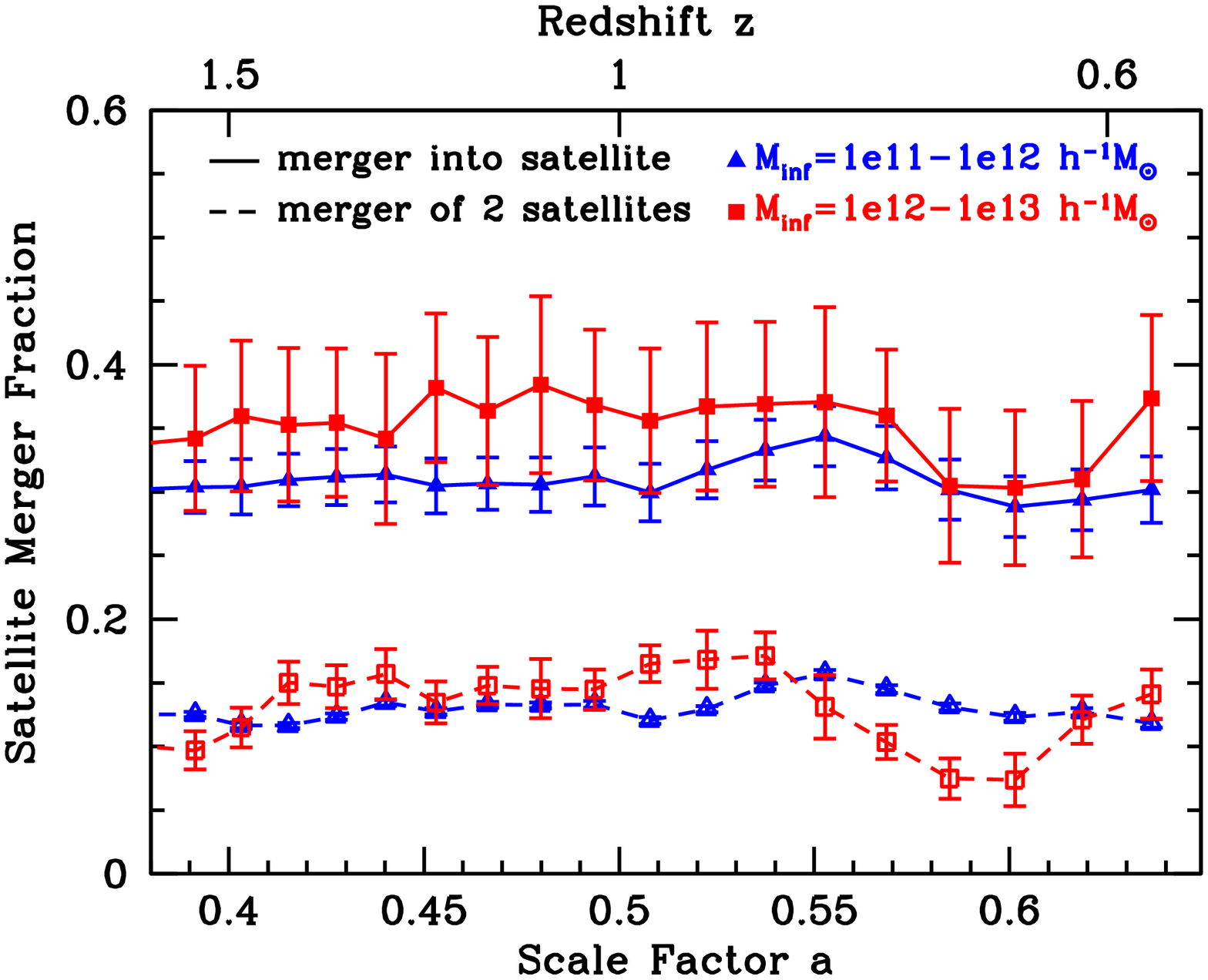}}
\end{center}
\vspace{-0.1in}
\caption{
Satellite merger fraction at $z=5-2.5$ (left) and $z=1.6-0.6$ (right) for 
$M_{\rm inf} = 10^{11}-10^{12} h^{-1}M_\odot$ (triangles) and 
$M_{\rm inf} = 10^{11}-10^{12} h^{-1}M_\odot$ (squares).
Closed point/solid lines indicate the fraction of all subhalo mergers that 
result in a satellite, while open points/dashed lines indicated the fraction of 
all subhalo mergers that arise from satellite-satellite parents.
} \label{fig:satmergefrac}
\end{figure*}
  
The stereotypical galaxy merger is a satellite coalescing with the central in 
its halo and producing a central merger remnant.
These mergers do dominate the merger population, both in parent types 
(central-satellite) and child type (central).
However, while satellites form the minority of the subhalo population at all 
epochs (see Fig.~\ref{fig:satfrac}), satellites are twice as likely to have had 
a recent merger as centrals of the same mass, regardless of mass and redshift.

Since the identities of satellite vs. central subhalos at this mass and 
redshift regime are not clear-cut (from switches), we characterize mergers both 
in terms of their parent types (central/satellite) and resulting child types.
For mergers resulting in centrals, this ambiguity is not important: 
$97\%$ arise from satellite-central parents, while the other $3\%$ arise from 
satellite-satellite parents during switches.\footnote{
A few percent arise from central-central parents, when the central regions of 
two halos coalesce so quickly that they are not seen as satellite-central 
subhalos given finite time resolution.
We include these as satellite-central parents.}

Recently merged satellites are a more varied population.
Figure~\ref{fig:satmergefrac} shows the contributions of satellite mergers to 
the overall merger populations as a function of the scale factor.\footnote{
Satellite merger fractions are boxcar-averaged across three consecutive outputs 
to reduce noise from small number statistics.}
The fraction of all subhalo mergers that result in a satellite is $\sim 30\%$  
for $M_{\rm inf}=10^{11}-10^{12} h^{-1}M_\odot$, with little dependence on 
redshift.
For $M_{\rm inf}=10^{12}-10^{13} h^{-1}M_\odot$ it is $15-20\%$ at $z \sim 2.5$ 
and rises to $\sim 35\%$ at $z<1.6$.
Thus, at lower redshift ($z<1.6$) where the satellite fraction asymptotes to 
$\sim 25\%$, the fraction of mergers that result in a satellite roughly 
reflects the satellite fraction as a whole.

Of these recently merged satellites, $20\%-35\%$ come from satellite-satellite 
parents within a single halo, while $\sim 7\%$ arise when a central-satellite 
merger occurs in a halo as it falls into a larger halo, becoming a satellite.
The rest arise from switches, i.e. a satellite merges with a central, and the 
resulting subhalo no longer is the most massive subhalo, thus becoming a 
satellite.
These switches occur primarily in halos only a few times more massive than the 
satellite, typically for satellites in close proximity to their central.
At higher halo masses, recently merged satellites are dominated by 
satellite-satellite parents.
These satellite-satellite mergers preferentially occur in the outer regions of 
a halo and are comparatively less common in the central regions.
We examine in more detail the environmental dependence of subhalo mergers in 
\citet{WetCohWhi09b}.

Considering instead only parent types, Fig.~\ref{fig:satmergefrac} shows that 
$5-10\%$ of all mergers come from satellite-satellite parents at $z \sim 2.5$, 
a fraction which increases to $10-15\%$ at $z=1.6$ and remains flat thereafter.
Approximately $80\%$ of all satellite-satellite mergers lead to a satellite
child, while the rest lead to a central during a switch.

%% MERGER COUNTS %%%%%%%%%%%%%%%%%%%%%%%%%%%%%%%%%%%%%%%%%%%%%%%%%%%%%%%%%%%%%%%
\section{Galaxy and Halo Merger Counts} \label{sec:counts}

\subsection{Counts of Recent Mergers}

\begin{figure}
\begin{center}
\resizebox{3in}{!}{\includegraphics{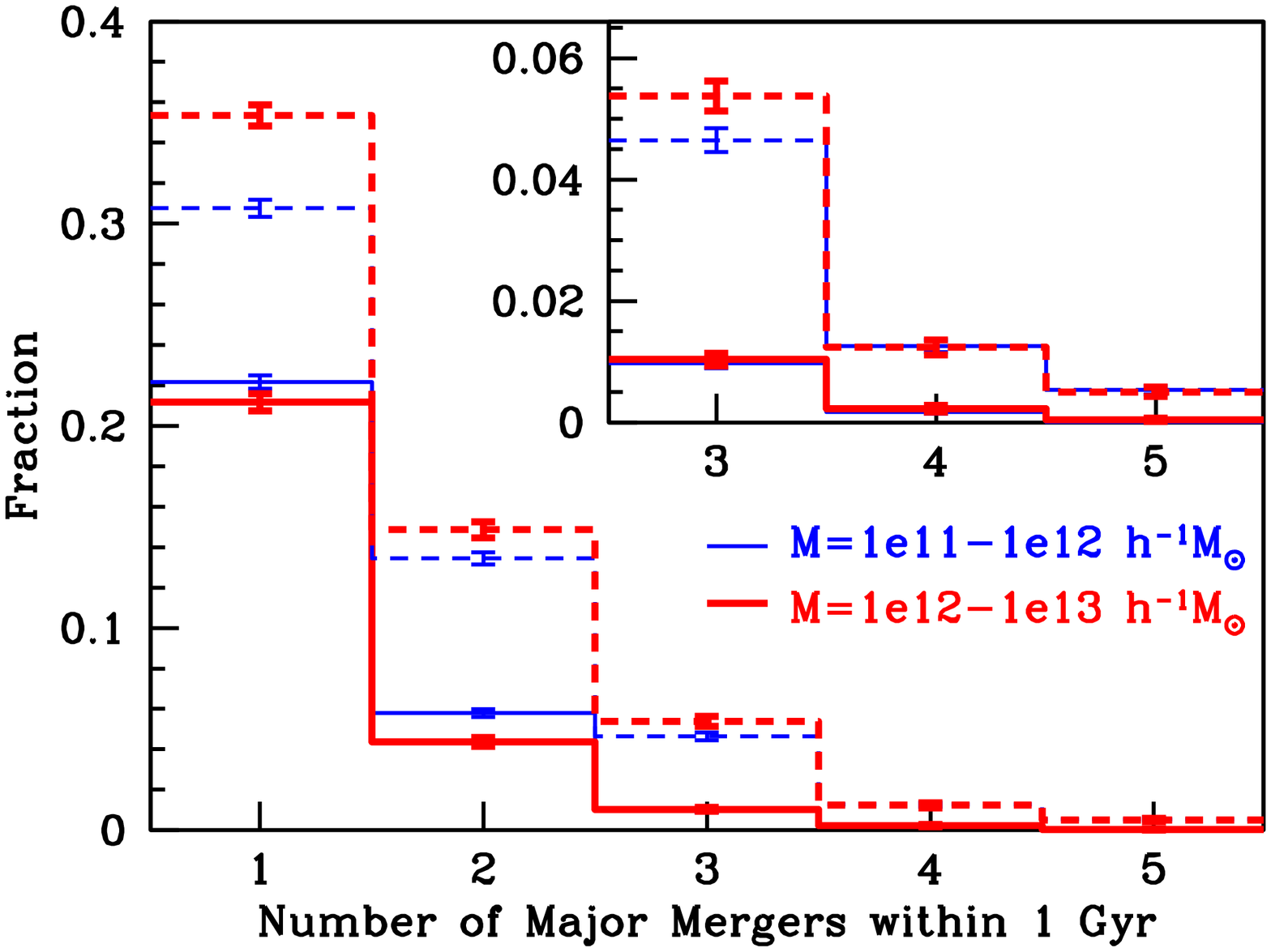}}
\resizebox{3in}{!}{\includegraphics{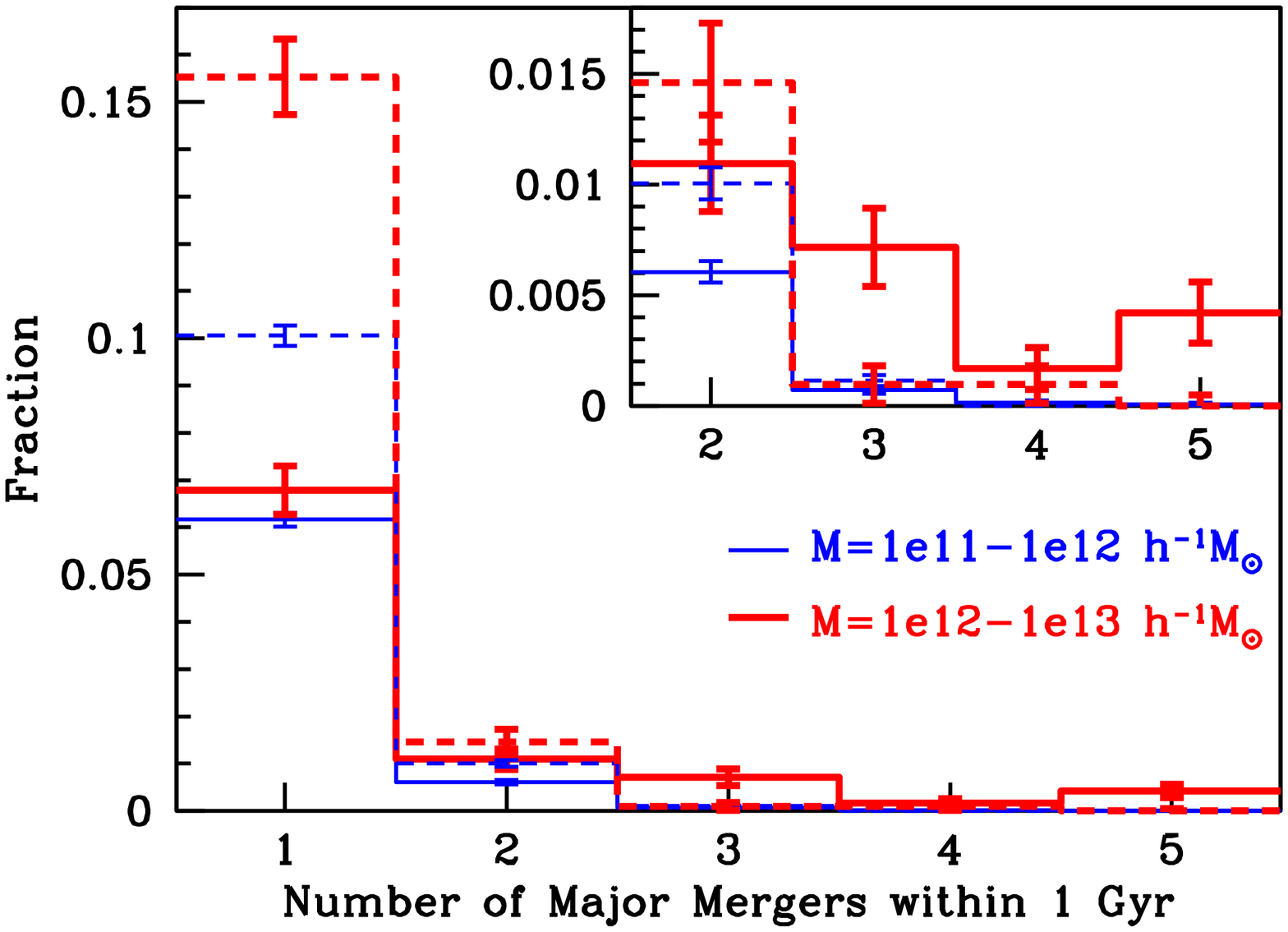}}
\end{center}
\vspace{-0.1in}
\caption{
\textbf{Top}: Fraction of subhalos at $z=2.6$ that have had a given number of 
mergers in the last $1$~Gyr, for 
$M_{\rm inf} = 10^{11}-10^{12} h^{-1}M_\odot$ (thin lines) and 
$M_{\rm inf} = 10^{12}-10^{13} h^{-1}M_\odot$ (thick lines).
Dashed lines show the same, but for host halos.
$30\%$ of subhalos have had at least one merger (independent of mass), while 
for halos this fraction is $50\%$ (lower mass) and $57\%$ (higher mass).
Inset shows detail for $>2$ mergers.
\textbf{Bottom}: Same, but at $z=1$.
$8\%$ of subhalos have had at least one merger (independent of mass), 
while for halos this fraction is $11\%$ (lower mass) and $17\%$ (higher mass).
High mass subhalos show a much larger fraction of multiple mergers.
Inset shows detail for $>1$ merger.
} \label{fig:nmerge}
\end{figure}

The distribution of the number of mergers per object within a fixed time 
interval gives the fraction of objects at a given epoch that might exhibit 
merger-related activity or morphological disturbance.\footnote{
This differs from the merger rates of \S\ref{sec:rates}, since we are tracking 
the histories of individual objects selected at a given redshift.}
Multiple mergers as well might contribute to specific properties, e.g., the 
formation and mass growth of elliptical galaxies \citep[e.g.,][]{BoyMaQua05,
Rob06,NaaKhoBur06,ConHoWhi07}.

Figure~\ref{fig:nmerge} shows the fraction of subhalos at $z=2.6$ and $z=1$ 
with a given number of mergers in the last $1$~Gyr.
At $z=2.6$, $30\%$ of subhalos have suffered at least one major merger.
Interestingly, this fraction is nearly constant across the mass regimes we 
probe.
In contrast, the fraction of halos with at least one major merger within 
$1$~Gyr is about twice as large, with stronger mass dependence: higher mass 
halos experience more mergers.
At high redshift, halos are also significantly more likely to have undergone
multiple mergers than subhalos, which builds up the satellite population.

At $z=1$, recent mergers of subhalos and halos become less common, with only 
$8\%$ of subhalos having suffered at least one major merger in the last $1$~Gyr.
Objects which have had 1 or 2 mergers are still more common for halos than 
subhalos, though high mass subhalos exhibit a much higher fraction of 3 or more 
mergers than halos of the same (infall) mass.
(Subhalos that have undergone 1 merger can be either satellites or centrals, 
those that have undergone 2 or more mergers are almost entirely centrals.)
The build-up of the satellite population at higher redshift has allowed massive 
centrals to experience multiple mergers at lower redshift.
This effect is stronger for more massive subhalos since they are more likely 
to be centrals, and they reside in higher mass halos with more massive 
satellites.

\subsection{Fraction ``On''}

\begin{figure*}
\begin{center}
\resizebox{3in}{!}{\includegraphics{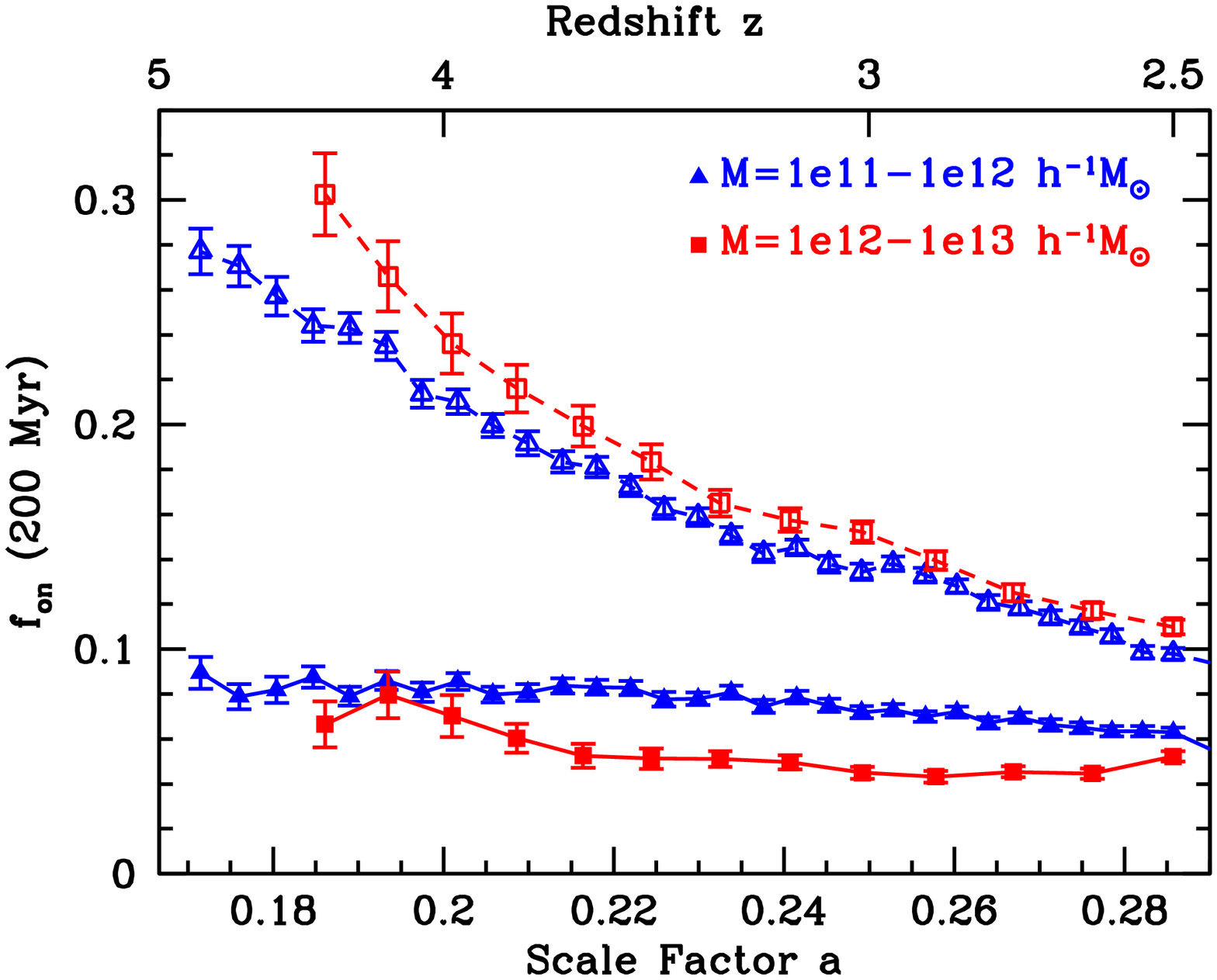}}
\resizebox{3in}{!}{\includegraphics{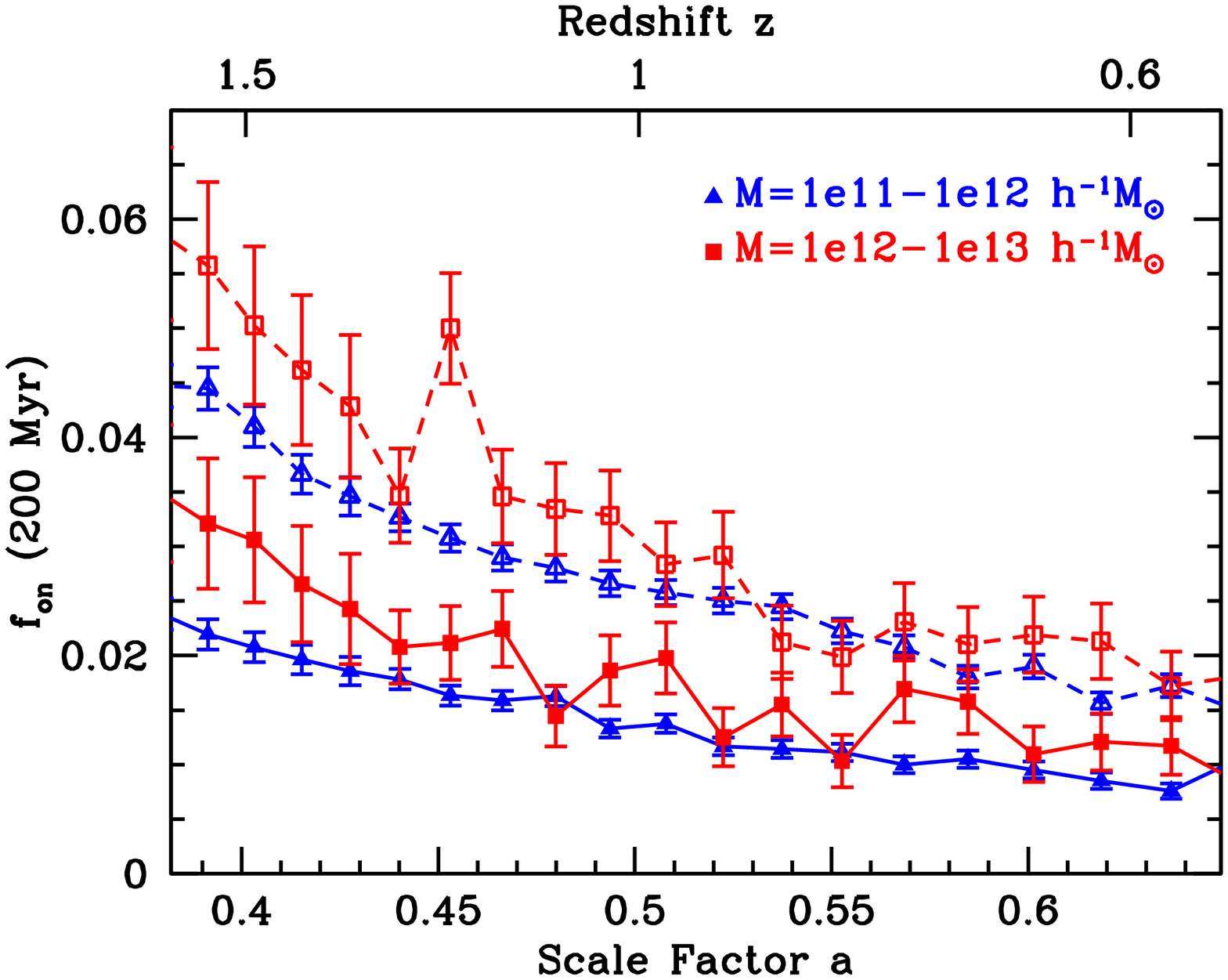}}
\end{center}
\vspace{-0.1in}
\caption{
\textbf{Left}: Fraction of halos that host a subhalo-subhalo merger within the 
last $200$~Myr (closed points/solid lines) at $z=5-2.5$ for 
$M_{\rm inf} = 10^{11}-10^{12} h^{-1}M_\odot$ (triangles) and 
$M_{\rm inf} = 10^{12}-10^{13} h^{-1}M_\odot$ (squares).
Open points/dashed lines show the fraction of halos of the same mass that have 
suffered a major halo-halo merger in the same time interval.
\textbf{Right}: Same, but at $z=1.6-0.6$.
} \label{fig:fon}
\end{figure*}

The fraction of halos that host recent mergers, $f_{\rm on}$, is of particular 
interest for quasar or starburst evolution models, and quasar/starburst 
feedback effects such as heating of the Intracluster Medium (ICM).
We select mergers up to $200$~Myr after coalescence, motivated by the expected 
time interval during which quasars or starbursts remain observable 
\citep[e.g.,][]{HopHerCox05}.
This time interval is only illustrative, though, as one expects relevant 
lifetimes to depend strongly upon galaxy mass and merger ratio.
Observables might depend upon dynamical time as well, although many quasar 
triggering effects might be related to microphysics--small scale interactions 
close to the merger--that do not evolve with time.\footnote{
If we scale our $f_{\rm on}$ time interval by the dynamical time, the slope of 
$f_{\rm on}$ becomes slightly shallower, but the qualitative results do not 
change.}

Figure~\ref{fig:fon} shows the evolution of $f_{\rm on}$ for halos hosting 
subhalo mergers within the last $200$~Myr.
The same quantity is shown for halos with recent mergers themselves.
At high redshift, $f_{\rm on}$ for halo mergers shows a steep decline from the
decreasing halo merger rate.
However, $f_{\rm on}$ for subhalo-subhalo mergers is flat from $z=2.5-5$, 
because the subhalo merger rate per object decreases while the number of 
massive satellites in a given mass halo rises, causing the number of massive 
subhalo mergers within the halo to remain constant.
At low redshift, where the satellite population grows more slowly, the 
evolution of this fraction for subhalo mergers more closely parallels that of 
halo mergers.

%%% EVOLUTION OF HALO OCCUPATION %%%%%%%%%%%%%%%%%%%%%%%%%%%%%%%%%%%%%%%%%%%%%%%
\section{Evolution of the Satellite Halo Occupation} \label{sec:hodevol}

The redshift evolution of the satellite galaxy populations of dark matter halos 
is shaped by halo vs. galaxy mergers: halo mergers create satellites while 
galaxy mergers remove them.\footnote{
Though if one applies a mass threshold to a population, this is not strictly 
true since mergers also scatter lower mass objects into the population.}
If the infall rate of satellites onto a halo is different than the satellite 
destruction rate, the satellite halo occupation will evolve with time.

As shown in Fig.~\ref{fig:mergerates}, at $z>2.5$, the merger rate of subhalos 
is significantly lower and shallower in slope than that of halos, implying that 
subhalos are being created faster than they are destroyed (at a rate decreasing 
with time).
Conversely, at $z<1.6$ the merger rates of halos and subhalos exhibit
approximately the same redshift dependence, and their amplitudes are similar
(also recall from \S\ref{sec:rates} that not all halo major merger lead to 
subhalo major mergers).
Thus, for halos of a fixed mass, we expect a rapid rise in the satellite halo 
occupation prior to $z \sim 2$ and a levelling-off with more gradual evolution 
at lower redshift.

\subsection{Satellite Halo Occupation in Simulation}

\begin{figure}
\begin{center}
\resizebox{3in}{!}{\includegraphics{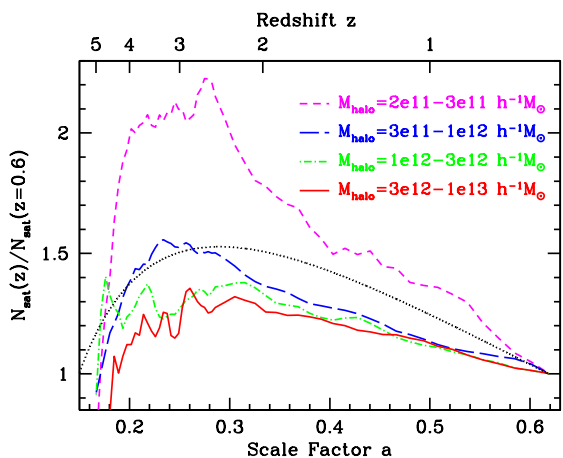}}
\end{center}
\vspace{-0.1in}
\caption{
Evolution of the average number of satellite subhalos per halo for satellites 
with $M_{\rm inf} > 10^{11} h^{-1}M_\odot$ and several halo mass bins.
Satellite counts are normalized to those at $z=0.6$ (the last output) and
boxcar averaged across 3 outputs to reduce the small number statistics noise at 
early times.
The dotted black line shows fit of Eq.~\ref{eq:hodevolfinal} for 3:1 mass ratio 
mergers using halo merger rate parameters from Table~\ref{tab:mergefits}.
All halos masses have a peak in satellite occupation at $z\sim2.5$, and similar 
trends persist for higher mass satellites.
Less massive halos exhibit stronger evolution with redshift, leading to a more
prominent peak.
} \label{fig:hodevol}
\end{figure}

The above trends are seen in Fig.~\ref{fig:hodevol}, which shows the evolution 
of the satellite occupation per halo, for satellites with 
$M_{\rm inf} > 10^{11} h^{-1}M_\odot$ in the $100 h^{-1}$~Mpc simulation.
Satellite occupation counts are normalized using the last output at $z=0.6$.
More massive satellites ($M_{\rm inf} > 10^{12} h^{-1}M_\odot$) have similar 
evolution, with a peak in the satellite halo occupation at $z \approx 2.5$.
For a fixed satellite infall mass, less massive halos exhibit stronger 
satellite occupation evolution with redshift, leading to a more prominent peak.

\subsection{Analytic Estimate of Satellite Halo Occupation}

The rate of change of the satellite subhalo population per halo, for a fixed 
satellite $M_{\rm inf}$, is given by the rate at which satellites fall into a 
halo (the halo merger rate) minus the rate at which satellites coalesce with 
the central subhalo
\begin{equation} \label{eq:hodevol1}
\frac{dN_{\rm sat}}{dt} = \frac{dN_{\rm halo,merge}}{dt}-\frac{dN_{\rm sat-cen,merge}}{dt} .
\end{equation}

The halo merger rate at all epochs is given by (\S\ref{sec:rates})
\begin{equation} \label{eq:hodevol2}
\frac{dN_{\rm halo,merge}}{dt} = A(1+z)^{\alpha} = Aa^{-\alpha}
\end{equation}
The timescale for the satellite to coalesce with its central subhalo after 
infall is given to good approximation by
\begin{equation}
t_{\rm sat-cen,merge} \approx \frac{C_o}{10}\frac{M_{\rm halo}/M_{\rm sat}}{\ln(1+M_{\rm halo}/M_{\rm sat})}t_{\rm Hubble} .
\end{equation}
where $C_o$ is a constant of order unity that accounts for the ensemble 
averaged satellite orbital parameters, and we leave it as our sole free 
parameter.
The rate of satellite destruction/coalescence is thus
\begin{equation} \label{eq:hodevol3}
\frac{dN_{\rm sat-cen,merge}}{dt} = \frac{10}{C_o}\frac{\ln(1+M_{\rm halo}/M_{\rm sat})}{M_{\rm halo}/M_{\rm sat}}H(z) .
\end{equation}
Combining Eqs.~\ref{eq:hodevol2} and \ref{eq:hodevol3} into 
Eq.~\ref{eq:hodevol1} one gets
\begin{equation} \label{eq:hodevol4}
\frac{dN_{\rm sat}}{dt} = Aa^{-\alpha} - \frac{10}{C_o}\frac{\ln(1+M_{\rm halo}/M_{\rm sat})}{M_{\rm halo}/M_{\rm sat}}H(z) .
\end{equation}

At high redshift, $H(z) \approx H_o(\Omega_{\rm m,o}a^{-3})^\frac{1}{2}$,
where $H_o \simeq 0.1 h$~Gyr$^{-1}$, which implies
\begin{equation} \label{eq:hodevolapprox}
\frac{dN_{\rm sat}}{dt} \approx Aa^{-\alpha} - \frac{1}{C_o}\frac{\ln(1+M_{\rm halo}/M_{\rm sat})}{M_{\rm halo}/M_{\rm sat}}\Omega_{\rm m,o}^\frac{1}{2}a^{-\frac{3}{2}} .
\end{equation}
Across all mass and redshift regimes, the power law index for halo mergers is 
$\alpha = 2-2.3$ (\S\ref{sec:rates}), and we find no additional dependence on 
merger mass ratio, suggesting a universal power law index.\footnote{
Though merger rates for more discrepant mass ratios have higher amplitudes, as 
given by Eq.~\ref{eq:mergeratio}.}
Thus, the creation and destruction terms in Eq.~\ref{eq:hodevolapprox} have 
differing dependencies on the scale factor, which implies they become equal at 
some redshift, where the satellite occupation per halo reaches a maximum.

Using the exact evolution of the Hubble parameter (in our cosmology) of 
$H(z) = H_o(\Omega_{\rm m,o}a^{-3}+\Omega_\Lambda)^{1/2}$ in 
Eq.~\ref{eq:hodevol4}, and integrating over $a$, the full evolution of the 
satellite occupation per halo is 
\begin{align}
  N_{\rm sat}(a) & =\frac{A}{H_o}\int da \frac{a^{-(\alpha+1)}}{H(z)} \notag \\
  & -\frac{10}{C_o}\frac{M_{\rm sat}}{M_{\rm halo}}\ln\left(1+\frac{M_{\rm halo}}{M_{\rm sat}}\right)\ln(a)+K
\label{eq:hodevolfinal}
\end{align}
where the constant $K$ accounts for initial/final conditions.
Figure~\ref{fig:hodevol} shows the resultant $N_{\rm sat}(a)$, normalized to 
the satellite occupation per halo at final output, using typical values for the 
halo merger rate from Table~\ref{tab:mergefits} ($A=0.032$ and $\alpha = 2.3$).
These values correspond to 3:1 mass ratio mergers, for satellites with 
$M_{\rm inf} = 10^{11} h^{-1}M_\odot$ this gives 
$M_{\rm halo} = 3\times10^{11} h^{-1}M_\odot$.
While the fit of this model to the simulation results is not exact, it nicely 
reproduces the general trends, especially given the simplicity of the model, 
which ignores halo/central mass growth, satellite-satellite mergers, and 
switches.
For instance, it correctly produces a lower peak in satellite occupation for 
halos with satellites of more discrepant mass ratios (more massive halos for a
fixed satellite mass, or less massive satellites for a fixed halo mass), 
relative to the amplitude at low redshift.
This is because the decreased infall times for smaller satellite-halo mass 
ratios cause a more dramatic fall (after the peak) in the satellite population 
for lower mass halos.

Agreement of this model with our simulations also requires $C_o \approx 2$, 
which can be compared with other work.
\citet{ZenBerBul05} and \citet{Jia08} find that the ensemble averaged 
satellite orbital circularity distribution is given by 
$\langle \epsilon \rangle = 0.5 \pm 0.2$, with no strong dependence on redshift
or satellite-halo mass ratio.
When applied to detailed dynamical friction timescale fits from simulation, 
this yields $C_o \approx 0.6$ \citep{BoyMaQua08} and 
$C_o \approx 1.4$ \citep{Jia08}.\footnote{
The $C_o$ values of these two fits agree only in extreme cases, i.e. maximally
circular orbits for \citet{BoyMaQua08} or maximally eccentric orbits for  
\citep{Jia08}.}
Taken at face value, our even higher value of $C_o$ means that our satellites 
are taking longer to merge, suggesting that the discrepancy does not arise from 
artificial over-merging in our simulations.
However, exact comparisons are difficult given the simple nature of our analytic
model, and because both of these groups use different halo and subhalo finding 
algorithms.
\citet{Jia08}, who used a simulation of roughly similar volume and mass 
resolution to ours, also incorporated hydrodynamics, which is likely to shorten 
the merger timescale since it introduces further dissipational effects to the 
subhalo orbits and reduces mass loss.
Compared with \citet{BoyMaQua08}, who performed much higher resolution 
simulations of isolated halo mergers, it is possible that our satellite 
subhalos experience more severe mass stripping upon infall, decreasing their 
mass and thus extending their subsequent infall time (see Eq.~\ref{eq:dynfric}).
A more detailed investigation of satellite infall timescales in a cosmological 
setting is needed, studies now in progress are targeting in particular the role 
of hydrodynamic effects \citep{Dol08,Sar08,Sim08}.

\subsection{Comparison to Other Work on Satellite Occupation Evolution}

Figure~\ref{fig:hodevol} shows a peak in the number of satellites per halo at 
$z \sim 2.5$.
Fundamentally, the reason for this peak is that we select subhalos of 
\textit{fixed} minimum \textit{infall} mass in halos of \textit{fixed} mass
across time.
If instead we examine the satellite occupation for halos above a minimum mass 
cut, the growth of massive halos (hosting more satellites) at late time would 
overwhelm the drop in the satellite population at a fixed halo mass, so the 
satellite halo occupation would grow monotonically and appear much like the 
satellite fraction in Fig.~\ref{fig:satfrac}, which ignores halo mass.\footnote{
Though the satellite halo occupation would have a higher amplitude since it 
measures $n_{\rm sat}/n_{\rm central}$ while the satellite fraction measures
$n_{\rm sat}/n_{\rm subhalo} = n_{\rm sat}/(n_{\rm central}+n_{\rm sat})$.}

These results agree with the interpretation that more massive halos have later 
formation times and longer satellite infall times, and thus host more 
substructure at a given epoch \citep{vdBTorGio05,ZenBerBul05}.
Similarly, for a fixed satellite infall mass, the satellite halo occupation 
evolves more rapidly for less massive halos, as \citet{ZenBerBul05} and 
\citet{DieKuhMad07} found, though their results were based upon subhalo 
instantaneous mass and maximum circular velocity (we compare these to 
$M_{\rm inf}$ in the Appendices).
  
However, the peak in satellite halo occupation in Fig.~\ref{fig:hodevol} does 
not appear in Halo Occupation Distribution (HOD) evolution studies by  
\citet{ZenBerBul05} because they use a fixed cut on instantaneous maximum 
circular velocity across time.
As shown in Appendix A, fixed $V_{\rm c,max}$ probes lower mass at higher 
redshift, and this evolution in satellite mass overwhelms the satellite 
evolution of Fig.~\ref{fig:hodevol}, leading to a monotonic rise in the 
satellite halo occupation with redshift.
Similarly, \citet{ConWecKra06} noted that the HOD shoulder (the halo mass where 
a halo hosts only a central galaxy) becomes shorter at higher redshift as an 
increasing fraction of low-mass halos host more than one galaxy, finding a 
monotonically increasing satellite population with redshift.
However, they compare fixed satellite number density (not mass) across 
redshift, which corresponds to a lower subhalo mass at higher redshift.
Again, this overwhelms the evolution of Fig.~\ref{fig:hodevol}, leading to a 
monotonic rise in the satellite population per halo with redshift.

In their semi-analytic model matched to simulation, \citet{vdBTorGio05} find 
that the average subhalo mass fraction of a halo always decreases with time, 
and they claim that, as a result, the timescale for subhalo mass loss 
(approximately the dynamical/infall time) is always smaller than the timescale 
of halo mass accretion (mergers).
This implies that the satellite infall rate is always higher than the halo 
merger rate, and so the satellite HOD always decreases with time.
However, this result refers to the total subhalo instantaneous mass per halo, 
not galaxy counts based on infall mass.
If instead we examine the evolution of the satellite occupation per halo as in  
Fig.~\ref{fig:hodevol} selecting the satellites based on \textit{instantaneous} 
subhalo mass instead of infall mass, we find a monotonic increase in the 
satellite occupation with no peak, in agreement with other authors above.

These examples all illustrate how the evolution of the HOD is dependent both 
on satellite mass assignment and selection of fixed mass vs. circular velocity 
vs. number density across time.

%% DISCUSSION %%%%%%%%%%%%%%%%%%%%%%%%%%%%%%%%%%%%%%%%%%%%%%%%%%%%%%%%%%%%%%%%%%
\section{Summary and Discussion} \label{sec:discussion}

Using high-resolution dark matter simulations in cosmological volumes, we have
measured the rates, counts, and types of subhalo (galaxy) major mergers at 
redshift $z=0.6-5$, describing their populations in terms of 
centrals/satellites and contrasting their merger properties with those of 
halos of the same (infall) mass.
We assign subhalos their mass at infall (with the capacity for mass growth
during satellite mergers), motivated by an expected correlation with galaxy 
stellar mass, but include no further semi-analytic galaxy modelling.
We select mergers requiring 3:1 or closer infall mass ratios, motivated by the 
expectation that these can trigger activity such as quasars, starbursts, and 
related objects such as Lyman break galaxies, submillimeter galaxies, and 
ULIRGs.
We highlight our main results as follows:

\begin{itemize}
\item The merger rate per object of subhalos is always lower than that of halos 
of the same (infall) mass.
Galaxies exhibit stronger mass dependence on the amplitude of their merger 
rate than halos, with more massive galaxies undergoing more mergers.
While the slope of the halo merger rate per object is essentially redshift 
independent, the slope of the galaxy merger rate is much shallower than that of 
halos at $z>2.5$ and parallels that of halos at $z<1.6$.
\item These differences in halos and subhalo merger rates arise because (1) 
halo mergers add mass to halos instantly, while central subhalo mass grows more 
gradually after a halo merger; (2) halo major mergers do not necessarily lead 
to subhalo major mergers, since central subhalos experience mass growth while a 
satellite subhalo's infall mass typically remains constant as it orbits; and 
(3) the satellite subhalo fraction grows with time, and satellites are twice as 
likely to be recent mergers as centrals of the same infall mass.
\item $15\%-35\%$ of all recently merged subhalos are satellites, though a 
significant fraction of these arise from satellite-central parents during 
switches.
$5\%-15\%$ of galaxy mergers arise from satellite-satellite parents, with a 
higher fraction at lower redshift.
\item At $z=2.6$~($z=1$), $30\%$~($8\%$) of galaxies have experience at least 
one major merger in the last $1$~Gyr, regardless of mass.
Halos are more likely to have experience multiple mergers in their recent 
history.
\item The likelihood of a halo to host a recently merged galaxy, $f_{\rm on}$,
does not evolve with time at $z>2.5$ and falls with time at $z<1.6$.
\item Comparing galaxy and halo merger rates allows one to understand the 
evolution of the satellite halo occupation, and we approximated this behaviour 
analytically including fits to our simulations.
Selecting subhalos on fixed infall mass, the satellite halo occupation for 
halos of a fixed mass increases with time at high redshift, peaks at 
$z\sim2.5$, and falls with time after that.
This implies similar evolution for the satellite galaxy component of the Halo 
Occupation Distribution.
\end{itemize}

Our results, based entirely on the dynamics of dark matter, represent an 
important but preliminary step towards quantifying the nature of galaxy mergers 
in hierarchical structure formation.
To compare to many observables we would need to include baryonic effects, and 
indeed such effects can provide corrections to the merger rates themselves 
\citep{Dol08,Jia08,Sar08}.
The merger rates here also do not include whether the subhalos are gas-rich 
(required for some observables) or not, though at the high redshifts we 
examine, we expect almost all galaxies to be gas-rich.
While satellites can be stripped of much of their gas before merging with their 
central \citep[e.g.,][]{Dol08,Sar08}, we have considered primarily massive 
satellites (relative to their host halos), which have short infall times and 
thus experience less gas stripping.

Timescales between observables and our measured merger event also play a role. 
In simulations, a quasar can appear up to $\sim 1$~Gyr after galaxy 
coalescence, though starbursts may occur more quickly 
\citep[e.g.,][]{HopHerCox05,SprDiMHer05b,Cox08}.
However, morphological disturbance is clearest during first passage and final 
coalescence \citep{LotJonCox08}.
The time scales for each signature to commence and/or persist also have a large 
scatter, ranging from $0.2$ to $1.2$~Gyr after the merger.
Finally, specific observations will also have specific selection functions.
The quantitative measurements provided here provide starting points for these 
analyses in addition to helping to understand the properties of galaxies and 
their mergers in general.

Unfortunately, our predictions are not easily compared to observations which 
estimate merger rates at $z \lesssim 1$ using close galaxy pairs or disturbed 
morphologies \citep[most recently,][]{Bel06,Kam07,Kar07,Lin08,LotDavFab08,McI08,
PatAtf08} since they have found that the close pair fraction evolves as 
$(1+z)^\alpha$ with a diverse range of exponents from $\alpha=0$ to $4$.
Furthermore, translating these observations into galaxy or halo merger rates 
requires including more physical effects, e.g. time dependent galaxy 
coalescence timescales \citep{KitWhi08,Mat08} or inclusion of changes in 
numbers of host halos with redshift \citep{BerBulBar06}.

While we were preparing this work for preparation, \citet{Sim08} appeared which 
considers detailed merger properties of satellite subhalos in a hydrodynamic 
simulation, and \citet{Ang08} appeared which considers satellite mergers in the 
Millennium simulation.
After this paper was submitted, \citet{SteBulBar08} appeared, which also 
considers merger rates as a function of mass, redshift, and mass ratio.

\section*{Acknowledgments}

We thank E. Scannapieco and K. Stewart for useful conversations and thank CCAPP 
at the Ohio State University for hospitality and hosting a meeting which 
started this project.
We also thank the referee for several useful suggestions.
A.W. gratefully acknowledges the support of an NSF Graduate Fellowship, J.D.C. 
support from NSF-AST-0810820 and DOE, and M.W. support from NASA and the DOE.
The simulations were analyzed at the National Energy Research Scientific 
Computing Center.

%% APPENDIX %%%%%%%%%%%%%%%%%%%%%%%%%%%%%%%%%%%%%%%%%%%%%%%%%%%%%%%%%%%%%%%%%%%%
\appendix
\section{Subhalo Mass and Circular Velocity}

Here we further elaborate on the details of subhalos in our simulations, 
including their radial density and circular velocity profiles and the relation 
between mass and maximal circular velocity, including its evolution with time.

\begin{figure*}
\begin{center}
\resizebox{3in}{!}{\includegraphics{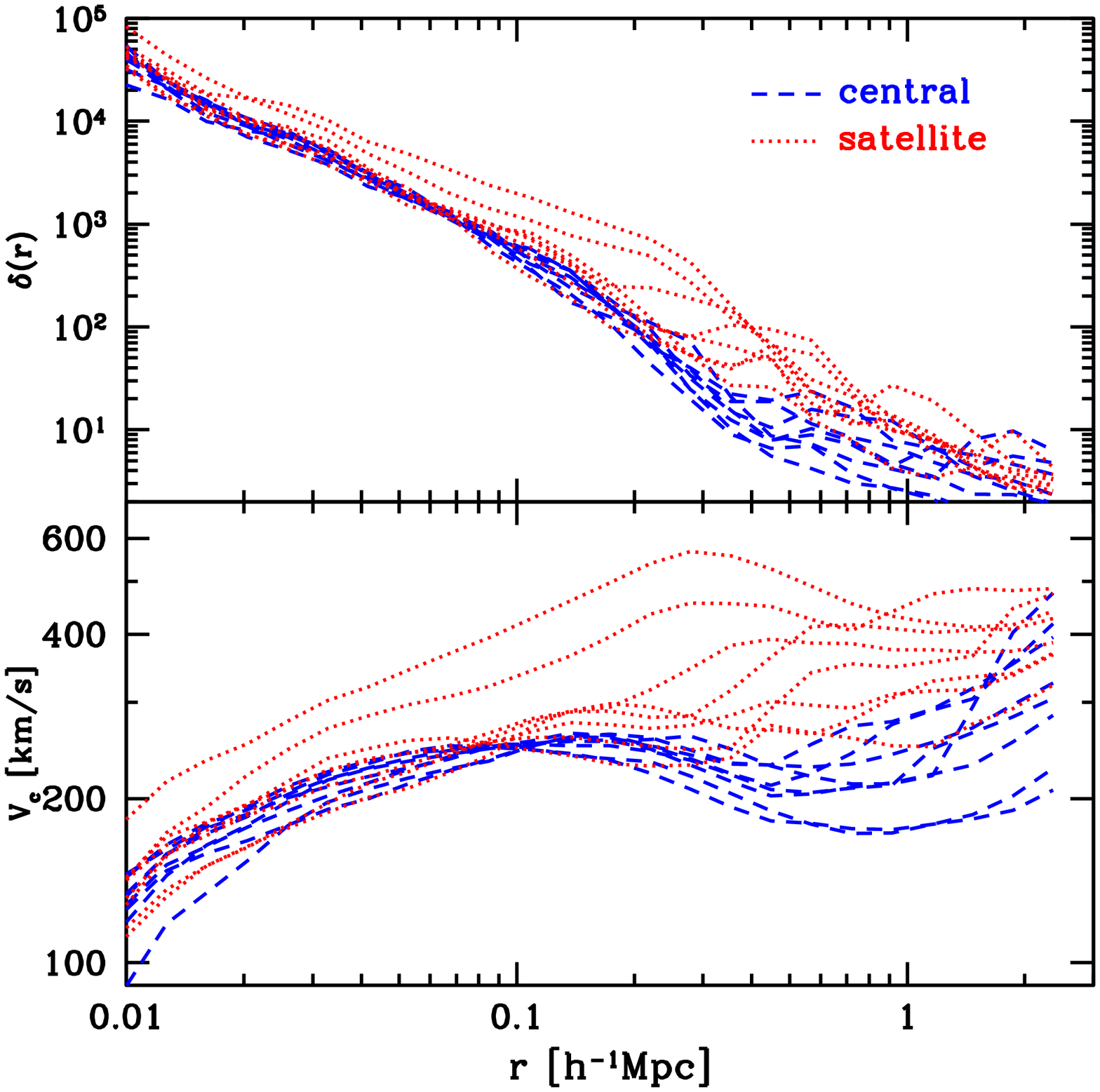}}
\resizebox{3in}{!}{\includegraphics{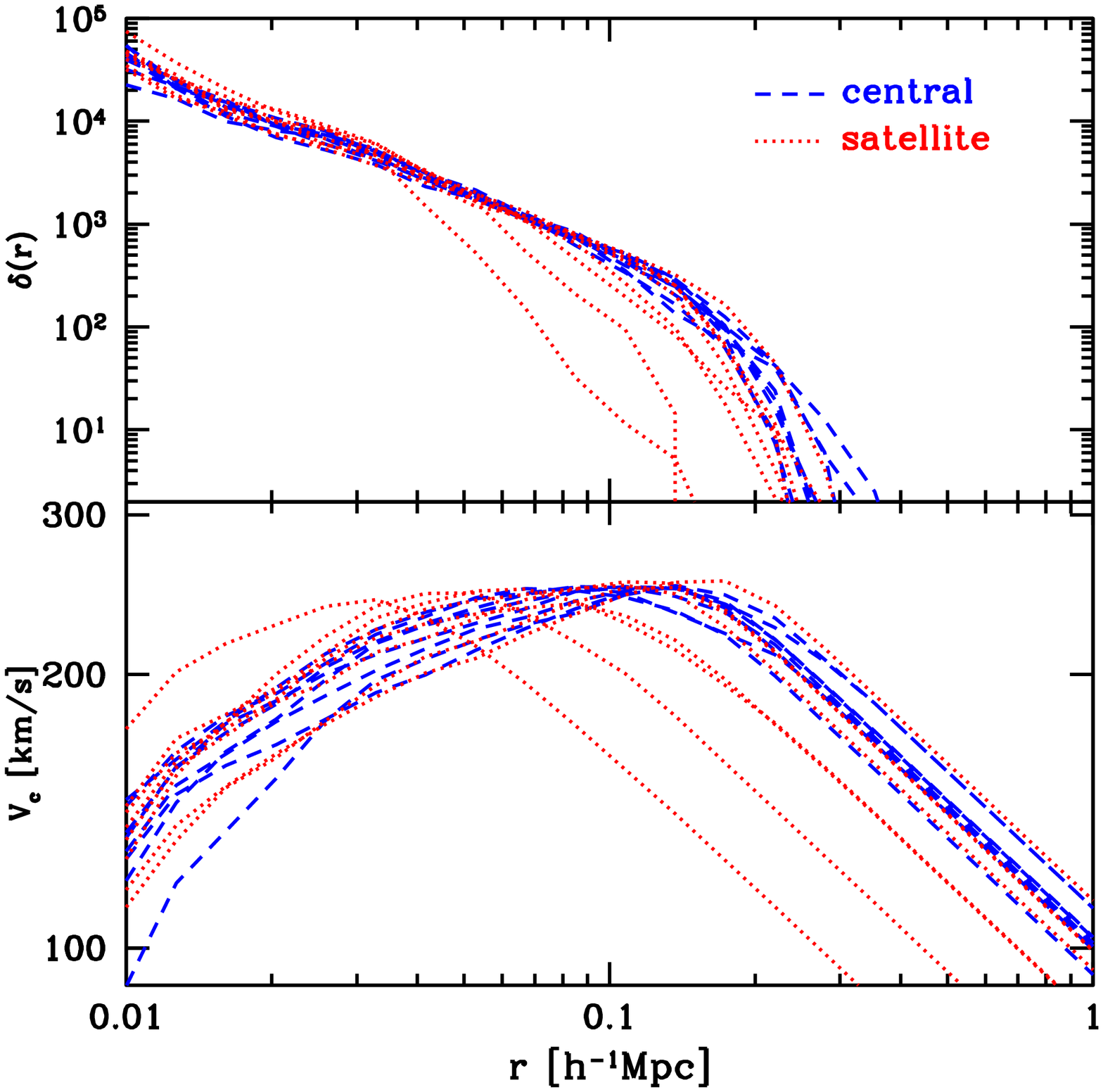}}
\end{center}
\vspace{-0.1in}
\caption{
\textbf{Left}: Radial density profiles (top) and circular velocity profiles 
(bottom) of all matter around 8 satellite (dotted) and 8 central (dashed) 
subhalos with $V_{\rm c,max} \simeq 250$~km/s at $z=2.6$.
\textbf{Right}: Same, but only for matter assigned to the subhalos.
} \label{fig:densityvcprofile}
\end{figure*}

Figure~\ref{fig:densityvcprofile} shows the radial density and circular 
velocity profiles for 8 satellite and 8 central subhalos with 
$V_{\rm c,max} \simeq 250$~km/s ($M \sim 10^{12} h^{-1}M_\odot$) at $z=2.6$ in 
the $100 h^{-1}$~Mpc simulation.
Circular velocity is defined as $V_c \equiv \sqrt{GM(<r)/r}$, and since 
subhalos follow NFW density profiles, with a break in the power-law density 
profile at the scale radius, $r_s$, they have a maximum value in their circular 
velocity profiles, $V_{\rm c,max}$, at $r_{\rm max} = 2.2r_s$.
The left panels show the radial profiles of all matter surrounding the 
subhalos, while the right panels show only that of matter assigned to the 
subhalos.
Satellites and centrals have similar profiles at small radii, though the right 
top panel shows that satellites exhibit signs of tidal truncation at 
$\sim 50 h^{-1}$~kpc.
This is also visible in the circular velocity profile in the bottom left panel, 
where $V_c$ for satellites rises sharply, exhibiting a transition to their host 
halos.
For the centrals, the rise in $V_c$ beyond $\sim 1 h^{-1}$~Mpc arises from 
neighbouring structures.

\begin{figure}
\begin{center}
\resizebox{3in}{!}{\includegraphics{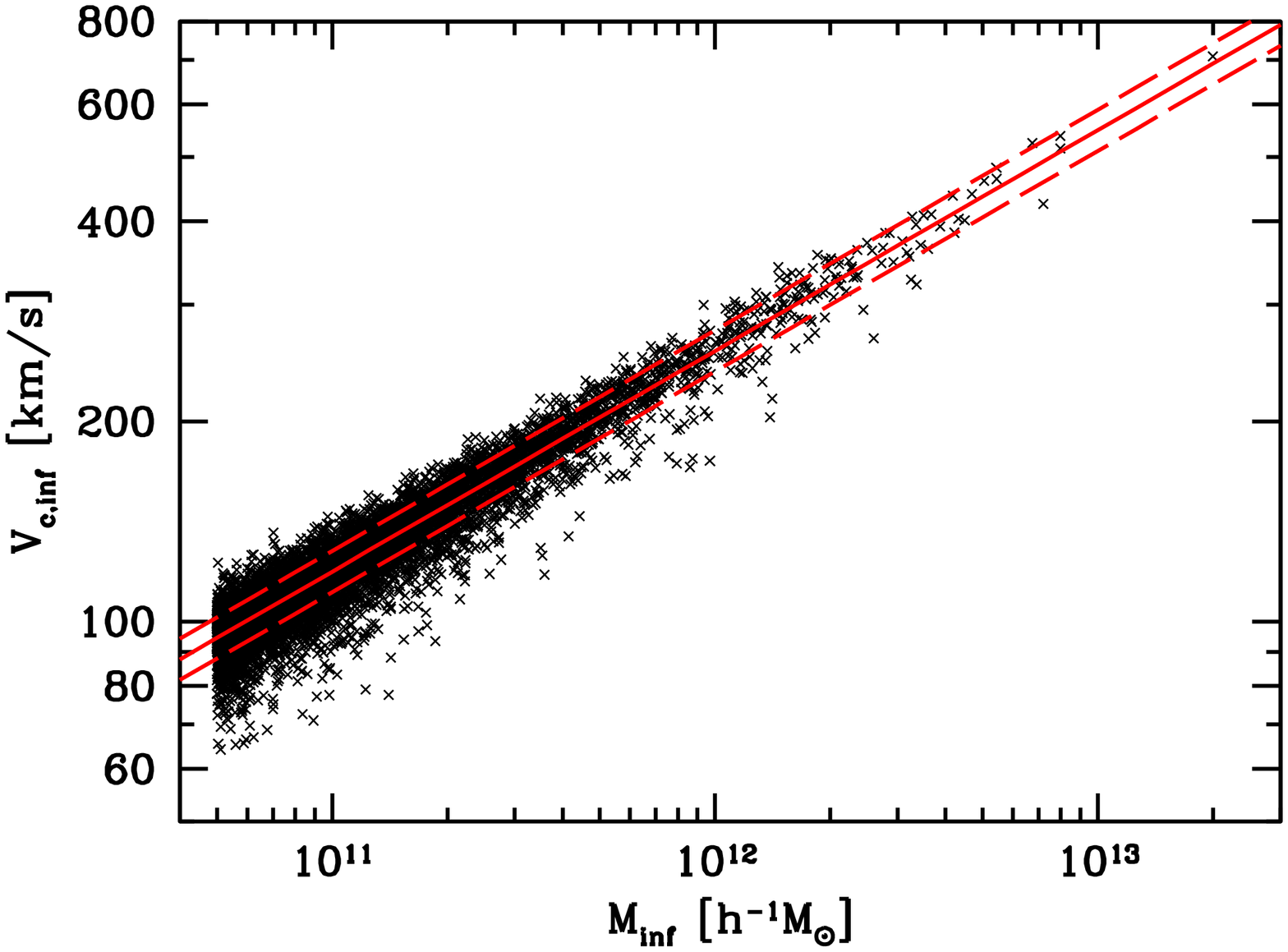}}
\end{center}
\vspace{-0.1in}
\caption{
Relation between subhalo infall maximum circular velocity, $V_{\rm c,inf}$, and 
subhalo infall mass, $M_{\rm inf}$, at $z=2.6$.
Black points show a $25\%$ sub-sample of all subhalos as a measure of scatter,
the solid red line shows the least squares fit to Eq.~\ref{eq:vcm}, and the 
dashed red lines show the $1\sigma$ scatter, for subhalos with 
$M_{\rm inf} > 10^{11} h^{-1}M_\odot$.
} \label{fig:vcinfminf}
\end{figure}

\begin{figure}
\begin{center}
\resizebox{3in}{!}{\includegraphics{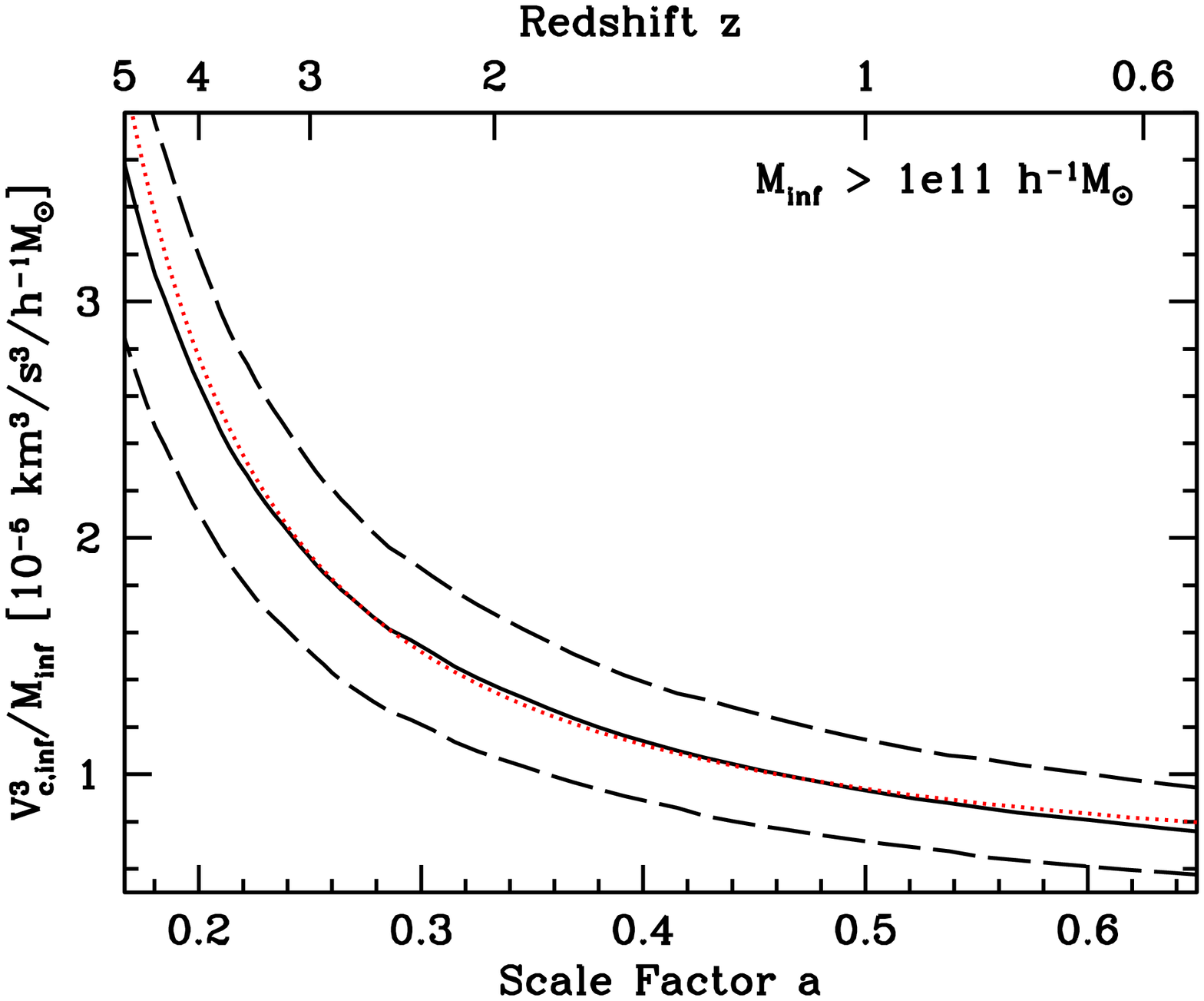}}
\end{center}
\vspace{-0.1in}
\caption{
Evolution of the ratio of subhalo infall maximum circular velocity to infall
mass, $B(z) = V_{\rm c,inf}^3/M_{\rm inf}$ (solid), and $1\sigma$ scatter (dashed) 
for subhalos with $M_{\rm inf} > 10^{11} h^{-1}M_\odot$.
Dotted line shows fit to $B(z)$ of Eq.~\ref{eq:vcmevol}.
} \label{fig:vcinfminfevol}
\end{figure}

Figure~\ref{fig:vcinfminf} shows the relation between subhalo infall mass, 
$M_{\rm inf}$, and infall maximum circular velocity, $V_{\rm c,inf}$, for 
subhalos at $z=2.6$.
We fit this relation to 
\begin{equation} \label{eq:vcm}
  V_{\rm c,inf}^{\gamma} = B(z) M_{\rm inf}
\end{equation}
for all subhalos above $10^{11} h^{-1}M_\odot$, finding ${\gamma} = 3$ holds to 
good approximation at all redshifts we examine, in agreement with the virial 
relation $V_{\rm c,max}^2 \propto M/R \propto M/M^{1/3} \propto M^{2/3}$.
The outliers with large $M_{\rm inf}$ relative to $V_{\rm c,inf}$ are 
satellites that experienced a major merger; under our prescription, a satellite 
child's $M_{\rm inf}$ is the sum of its parents' $M_{\rm inf}$, but a 
child's $V_{\rm c,inf}$ is that of its highest $V_{\rm c,inf}$ parent.

Fixing ${\gamma} = 3$, Fig.~\ref{fig:vcinfminfevol} shows the evolution of the 
amplitude $B(z) \equiv V_{\rm c,inf}^3/M_{\rm inf}$.
A subhalo of a given mass has a higher maximum circular velocity at higher 
redshift, reflective of the increased density of the universe when the subhalo 
formed.
As a subhalo subsequently accretes mass, its $V_{\rm c.max}^3$ grows more slowly 
than it mass (in cases of slow mass growth, we find $V_{\rm c.max}^3$ can remain 
constant).
This is in agreement with \citet{DieKuhMad07}, who found that halos undergoing 
mild mass growth (no major mergers) had less than $10\%$ change in 
$V_{\rm c,max}$ and $r_{\rm max}$.
We find that the evolution of $B(z)$ (within the redshifts we probe) can be 
well-approximated by 
\begin{equation} \label{eq:vcmevol}
B(z) = 6.56\times10^{-6} e^{0.36z} (\mbox{km/s})^3 h M_\odot^{-1} .
\end{equation}
Thus, $M_{\rm inf} = 10^{11}~(10^{12}) h^{-1}M_\odot$ subhalos correspond to 
$V_{\rm c,inf} \simeq 120~(250)$~km/s at $z=2.6$ and 
$V_{\rm c,inf} \simeq 100~(200)$~km/s at $z=1$.
Conversely, for fixed $V_{\rm c,inf}$, a subhalo is about half as massive at 
$z=2.6$ than at $z=1$.

Since we fit relations of satellite subhalo properties at infall, our results 
above are applicable equally to satellite and central subhalos.
In addition, our results change by only a few percent if we instead consider 
host halos.

%% SATELLITE POPULATION %%%%%%%%%%%%%%%%%%%%%%%%%%%%%%%%%%%%%%%%%%%%%%%%%%%%%%%%
\section{Satellite Subhalo Mass Function} \label{sec:satmassfunction}

\begin{figure}
\begin{center}
\resizebox{3in}{!}{\includegraphics{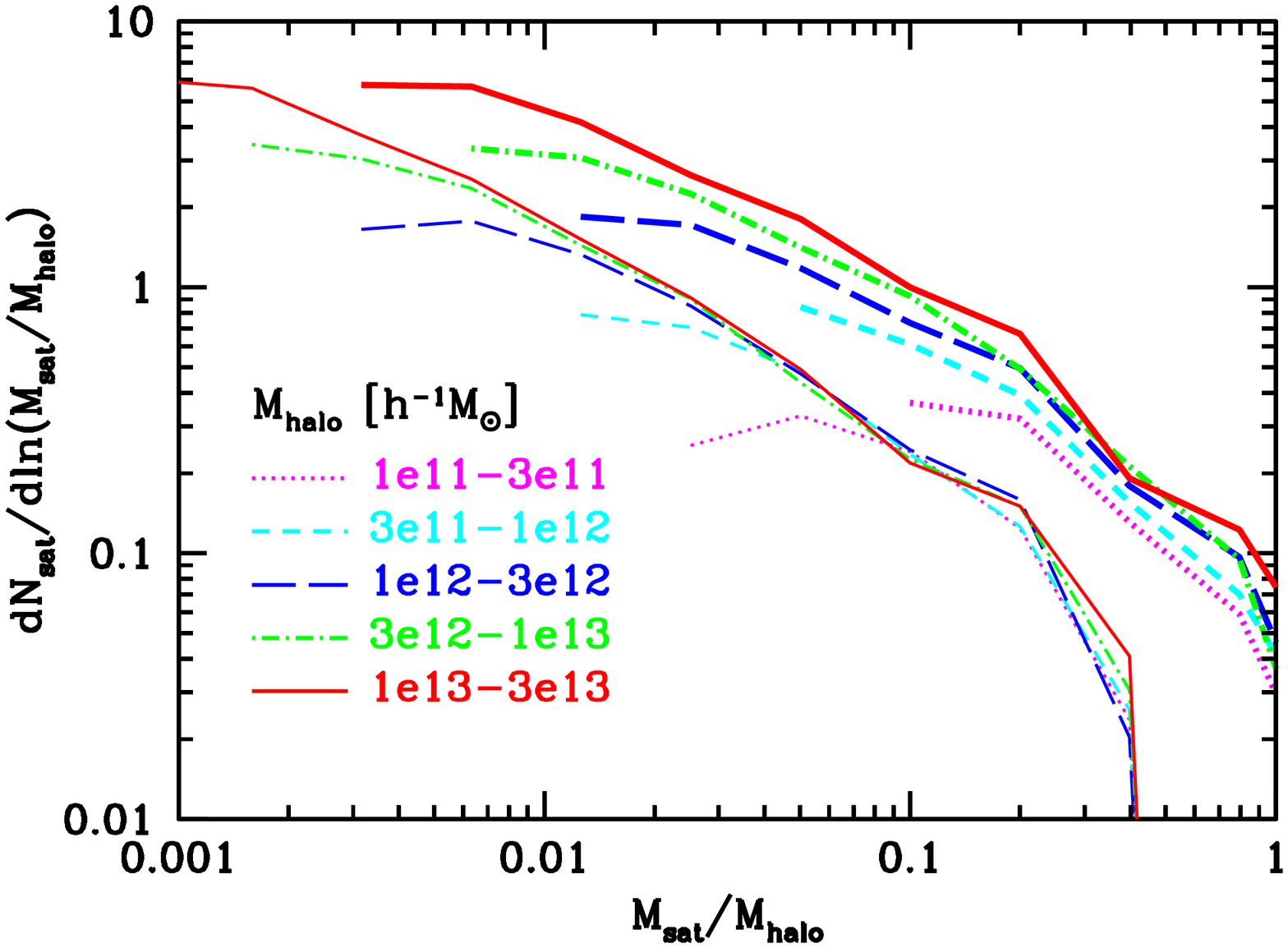}}
\end{center}
\vspace{-0.1in}
\caption{
Satellite subhalo (scaled) mass function, for various host halo mass bins, at 
$z=1$.
Thick curves show satellite mass selected on $M_{\rm inf}$, while thin curves 
show satellite mass selected on instantaneous bound mass.
While the instantaneous bound mass function exhibits no dependence on halo 
mass, the infall mass function has a higher amplitude for more massive halos.
An appreciable number of satellites exists at 
$M_{\rm sat,inf}/M_{\rm halo} \approx 1$ because of switches.
Below the rollover at high satellite mass, both mass functions scale as 
$\frac{dN_{\rm sat}}{d\ln(M_{\rm sat}/M_{\rm halo})} \propto M_{\rm sat}^{-0.9}$ 
in agreement with instantaneous satellite subhalo bound mass functions found by 
numerous authors (see text).
} \label{fig:satmassfunction}
\end{figure}

Figure~\ref{fig:satmassfunction} shows the satellite subhalo (scaled) mass 
function vs. the ratio of satellite mass to halo mass, for various host halo 
mass bins at $z=1$.
Thick curves show satellite masses selected on $M_{\rm inf}$, while thin curves 
show satellite masses selected on instantaneous bound mass.
The rollover in the $M_{\rm inf}$ curves at low satellite mass indicates where 
satellites become numerically disrupted by resolution effects.
Numerical disruption occurs at higher satellite $M_{\rm inf}$ for more massive 
halos, indicating that satellites of a fixed $M_{\rm inf}$ experience more 
pronounced tidal stripping in higher mass halos, where dynamical friction 
timescales and central densities are higher.
The rollover in the highest halo mass bin occurs at 
$M_{\rm sat,inf} \approx 10^{11} h^{-1}M_\odot$, which sets our minimum subhalo
mass for robust tracking.

We find that the scaled instantaneous bound mass function exhibits little-to-no 
systematic dependence on halo mass, in agreement with \citet{Ang08}.\footnote{
We thank the referee for suggesting we show this quantity.}
This is in contrast to the scaled infall mass function, which shows more 
satellites at a given mass ratio for more massive halos.
This difference is driven by subhalo mass stripping.
Averaged over the entire satellite population at this redshift, the satellite 
instantaneous bound mass is $\sim 30\%$ that of $M_{\rm inf}$, as can be seen by 
the x-axis offset of the solid and dashed curves.
However, satellites exhibit less average mass loss in low-mass halos than 
high-mass halos, the instantaneous to infall mass ratios being $40\%$ and 
$25\%$, respectively.
This is considerably higher than the $5\%-10\%$ at $z=1$ found in the 
semi-analytic model of \citet{vdBTorGio05}.
Additionally \citet{vdBTorGio05} and \citet{GioTorvdB08} found the opposite 
trend with halo mass, i.e., that the average mass loss of satellites is higher 
for lower mass halos.
They find that this arises because lower mass halos form (and accrete their 
subhalos) earlier, when the dynamical/stripping timescale is shorter.
Thus, the satellites of lower mass halo are stripped both more rapidly and over 
a longer time period \citep[see also][]{ZenBerBul05}.
Finally, \citet{GioTorvdB08} find that the scaled mass functions of subhalos at 
infall does not depend on halo mass, in seeming contrast with 
Fig.~\ref{fig:satmassfunction}.

These discrepancies likely arises because we examine the masses of extant 
subhalos in our simulation, while \citet{vdBTorGio05} and \citet{GioTorvdB08} 
track subhalo mass loss much longer than our simulation does.
Their semi-analytical model of subhalo mass loss has no prescription for 
central-satellite mergers, which preferentially serve to reduce highly stripped 
satellites from our sample.\footnote{
See \citet{TayBab05} and \citet{ZenBerBul05} for detailed comparisons of 
simulated subhalos with analytic models.}
Since satellites of a given infall mass to halo mass ratio have lower mass in 
lower mass halos, they are closer to our minimum subhalo finding mass threshold.
Thus, a fixed amount of stripping will cause lower mass halos to have a reduced 
population of satellites of a given infall mass to halo mass ratio.
It is unclear at what level of subhalo mass stripping we should expect the 
galaxies they host to become disrupted as well.

Various studies using high resolution simulations have explored in detail the 
slope of the subhalo mass function \citep{DeL04,GaoDelWhi04,DieKuhMad07,
MadDieKuh08,Ang08}, finding that the (instantaneous) mass function of subhalos 
goes as $N_{\rm sat}(>M_{\rm sat}) \propto (M_{\rm halo}/M_{\rm sat})^{\alpha}$, 
with $\alpha = 0.9-1$.\footnote{
$N_{\rm sat}(>M_{\rm sat})$ has the same power law dependence on subhalo mass 
as $\frac{dN_{\rm sat}}{d\ln(M_{\rm sat}/M_{\rm halo}))}$.}
We note, though, that these fits are based on cuts on instantaneous subhalo 
mass or circular velocity, not those at infall.
However, we find that, between the rollover in $M_{\rm inf}$ at low and high 
satellite mass, the slopes of the mass functions selected on infall and 
instantaneous bound mass are the same within error, and $\alpha = 0.9$ fits our 
mass function selected on either mass.
The similarity of the two slopes also implies that the mass loss rate of 
satellites does not depend strongly on satellite mass, in agreement with 
\citet{GioTorvdB08}.

%% BIBLIOGRAPHY %%%%%%%%%%%%%%%%%%%%%%%%%%%%%%%%%%%%%%%%%%%%%%%%%%%%%%%%%%%%%%%%
\bibliography{ms.bbl}

\label{lastpage}

\end{document}